\documentclass[a4paper,11pt]{article}
\pdfoutput=1 

\usepackage{jheppub} 
\usepackage[T1]{fontenc} 
\usepackage{amsmath}
\usepackage{color}

\def\l{\left(}
\def\r{\right)}

\makeatletter

\def\lsim{\compoundrel<\over\sim}
\def\compoundrel#1\over#2{\mathpalette\compoundreL{{#1}\over{#2}}}
\def\compoundreL#1#2{\compoundREL#1#2}
\def\compoundREL#1#2\over#3{\mathrel
         {\vcenter{\hbox{$\m@th\buildrel{#1#2}\over{#1#3}$}}}}
\makeatother

\title{Sgoldstino signal at FASER: \\ prospects in searches for supersymmetry}

\author{Sergey Demidov,}
\author{Dmitry Gorbunov}
\author{and Dmitry Kalashnikov}
\affiliation{Institute for Nuclear Research of the 
Russian Academy of Sciences, Moscow 117312, Russia \\and\\
Moscow Institute of Physics and Technology, Dolgoprudny 141700, Russia}

\emailAdd{demidov@ms2.inr.ac.ru}
\emailAdd{gorby@ms2.inr.ac.ru}
\emailAdd{kalashnikov.d@phystech.edu}

\abstract{We investigate FASER@LHC perspectives in searches for light ($0.1-5$\,GeV)  sgoldstinos in models with low energy ($10-10^4$\,TeV) supersymmetry breaking. We consider flavor conserving and flavor violating couplings of sgoldstinos to Standard Model fermions and find the both options to be testable at FASER. Even the first FASER run allows one to probe interesting patches in the model parameter space, while the second run, FASER-II, with significantly larger detector fiducial volume, gives a possibility to thoroughly explore a wide class of supersymmetric extensions of particle physics complementary to those probed at LHC with ATLAS and CMS detectors.}

\usepackage{graphicx}
\graphicspath{{/}}	
\DeclareGraphicsExtensions{.pdf,.png,.jpg}	
\begin{document}

\maketitle
\flushbottom

\section{Introduction}
\label{Sec:Intro}

Supersymmetry provides with a technically natural solution to the gauge hierarchy problem\,\cite{Giudice:2013yca}, and despite the discouraging absence of superpartners of the Standard Model (SM) particles at the energy scale available for investigation with Large Hadron Collider (LHC), it remains the most developed extension of the modern particle physics. Additional supports follow from the String theory (theoretical motivation based on unification of all the fundamental interactions\,\cite{Schwarz:1982jn,Witten:1985xc}) and cosmology (the lightest superpartner as a Dark Matter candidate\,\cite{Jungman:1995df,Bertone:2004pz}). The theoretical schemes and cosmological applications  may well be consistent with the negative results of LHC, e.g. recall the idea of split supersymmetry\,\cite{Arkani-Hamed:2004ymt,Giudice:2004tc}. At the same time, the phenomenology may differ from the standard expectations, and the first evidence for supersymmetry may be associated with new particles from hidden sectors rather than with SM superpartners, which can be sufficiently heavy beyond the threshold of direct production at LHC. 

This situation can be generically realized in scenarios where supersymmetry is broken at low energy scale, which can be achieved in models with no-scale supergravity\,\cite{Ellis:1984kd,Ellis:1984xe} or gauge-mediation\,\cite{Giudice:1998bp,Dubovsky:1999xc}. While the structure of hidden sectors is model-dependent, there must be a sector, where supersymmetry gets spontaneously broken, and it contains  the Goldstone supermultiplet at least. 
Components of the latter can be sufficiently light, below the LHC threshold, and its couplings to the SM fields, suppressed by powers of the supersymmetry breaking scale, see e.g.\,\cite{Gorbunov:2001pd,Brignole_2003cm}, are moderately feeble, giving a chance to probe this new physics with large enough statistics of proton collisions. 

Components of the Goldstone supermultiplet with masses of tens of GeV and heavier most effectively can be searched for at CMS and ATLAS, see e.g. Refs.\,\cite{Perazzi:2000ty,Gorbunov:2000ht,Gorbunov:2002er,Demidov:2020jne}, while lighter particles, especially with flavor-violating couplings to quarks are most convenient for LHCb, see e.g. Refs.\,\cite{Gorbunov:2000th,Gorbunov:2000cz,Demidov:2011rd} for sets of predictions and Refs.\,\cite{LHCb:2012myk,LHCb:2013zjg,LHCb:2017rdd,LHCb:2021iwr} for the negative tests. More options for light particles open up with construction of new forward detectors, especially those placed at a noticeable distance from the proton collision point. Indeed, these particles propagate mostly along the beam line because of  kinematics. At the same time the forward flux of mesons produced in the same collisions gets suppressed at large distance due to scattering off matter and weak decays, which allows one to distinguish  particle signals in a specially designed far detector. 

In this paper we consider recently proposed\,\cite{Feng_2018,Ariga_2019} and approved experiment FASER\,\cite{fasercollaboration2019faser} as a tool to investigate supersymmetric extensions of the SM with light {\it sgoldstinos}\,--\,scalar and pseudoscalar components of the Goldstone supermultiplet. Sgoldstinos are $R$-even and can decay into a pair of the SM particles providing with a clear signal signature at FASER. They can be produced directly in proton collisions, in scattering of secondary photons (the Primakoff effect) and as well as in decays of mesons which appear in hadronic showers.  While sgoldstino masses depend on details of the sector where supersymmetry is broken, their coupling constants to the SM particles are determined by ratios of the supersymmetry breaking parameters~\cite{Brignole,Brignole_2003cm}, e.g. superpartner masses, and the supersymmetry breaking scale.    Hence, the measurement of sgoldstino couplings allows one to infer the information about the superpartner hierarchy and the hidden-sector dynamics responsible for the supersymmetry breaking in the entire model. 

{ The FASER experiment has been suggested as a specially designed tool to search for light exotic feebly interacting massive particles produced in proton-proton scatterings, and the supersymmetric models with sgoldstinos provide with one of such examples. Many others are considered in \cite{fasercollaboration2019faser,Feng_2018,Ariga_2019}, for recent ideas see e.g. \cite{Li:2021tsy,Araki:2020wkq,Okada:2019opp} and detailed review\,\cite{Feng:2022inv}.  }

The paper is organized as follows. Sgoldstino effective lagrangian, sgoldstino decay modes and meson decays into sgoldstinos are considered in Sec.\,\ref{Sec:Sgoldstino}. The FASER experiment is briefly described in Sec.\,\ref{Sec:FASER}. Both direct and indirect production of light sgoldstinos at LHC and the corresponding sgoldstino flux through the FASER detector are discussed in Secs.\,\ref{Subsec:direct} and\,\ref{Subsec:indirect}. The sgoldstino signal is investigated in Sec.\,\ref{Sec:sensitivity}, where we present the expected FASER sensitivity to parameters of the supersymmetric models with light sgoldstinos. Both flavor-conserving and flavor-violating sgoldstino couplings to the SM fermions are studied, and both stages, FASER-I and FASER-II are explored. We summarize the obtained results and discuss the questions which deserve further investigations in Sec.\,\ref{Sec:conclusion}.

\section{Light sgoldstinos}
\label{Sec:Sgoldstino}

If supersymmetry exists in  nature it must be spontaneously broken. In a wide class of models the supersymmetry breaking happens in a hidden sector when auxiliary component of some chiral supermultiplet gets non-zero vacuum expectation value $F\neq0$. Naturally this supermultiplet is called {\it Goldstone}, as it contains the Goldstone field, massless fermion, {\it goldstino}. The latter becomes a longitudinal component of gravitino in models with local supersymmetry (supergravity). Goldstino superpartners, scalar $S$ and pseudoscalar $P$ {\it sgoldstinos}, remain massless at tree level, and gain their masses, $m_S$ and $m_P$, due to quantum corrections. The scale of sgoldstino masses depends on details of the hidden sector. We consider this scale to be below 10\,GeV in what follows. We also assume all  the superpartners  of  the SM particles and particles from the hidden sector (except for gravitino), to be much heavier, and neglect their possible direct impact on sgoldstino phenomenology. Likewise we neglect sgoldstino interactions with gravitino. The latter is the $R$-odd lightest superpartner, and so is naturally stable or very-long lived (if $R$-parity is slightly broken). Sgoldstino can decay into a pair of gravitinos, however corresponding partial decay width is suppressed with respect to the others by a factor $\left(\frac{m_{S,P}^2}{m_{soft}^2}\right)^2$, where $m_{soft}$ is a common scale of soft supersymmetry breaking parameters\,\cite{Gorbunov:2000th}. As we consider models with very light sgoldstinos, their decays to gravitinos can be safely ignored.

\subsection{Sgoldstino interaction lagrangian}
\label{Subsec:lagr}

Sgoldstino interaction with the SM particles can be derived adopting the spurion technique\,\cite{Brignole} for the Goldstone superfield, see e.g.\,\cite{Gorbunov:2001pd}. As we are interested in light scalar $S$ and pseudoscalar $P$ sgoldstinos, the relevant for our study couplings are those to photons (field strength tensor $F_{\mu\nu}$) and gluons (field strength tensor $G_{\mu\nu}^a$, $a=1,\dots,8$) \,\cite{Gorbunov:2000th}, 
\begin{eqnarray}
    \label{Lagr-photons}
    {\cal L} = &-\frac{M_{\gamma\gamma}}{2\sqrt{2}F}SF_{\mu\nu} F^{\mu\nu} 
    +\frac{M_{\gamma\gamma}}{4\sqrt{2}F}PF_{\mu\nu} F_{\lambda\rho}\,\epsilon^{\mu\nu\lambda\rho} \\
    \label{Lagr-gluons}
    &-\frac{M_3}{2\sqrt{2}F}SG^a_{\mu\nu} G^{\mu\nu\;a} 
    +\frac{M_3}{4\sqrt{2}F}PG^a_{\mu\nu} G^a_{\lambda\rho}\,\epsilon^{\mu\nu\lambda\rho}
\end{eqnarray}
as well as to quarks $q_i$, $i=1,\dots,6$ and charged leptons $l_i$, $i=1,2,3$,  
\begin{equation}
    \label{Lagr-quarks}
    {\cal L} = -S\,\frac{v A_Q y^q_{ij}}{\sqrt{2}\,F}\bar q_i q_j 
    -P\,\frac{v A_Q y^q_{ij}}{\sqrt{2}\,F}\bar q_i \gamma_5 q_j  
   -S\,\frac{v A_l y^l_{ij}}{\sqrt{2}\,F}\bar l_i l_j 
    -P\,\frac{v A_l y^l_{ij}}{\sqrt{2}\,F}\bar l_i \gamma_5 l_j \,.  
\end{equation}
Here the supersymmetry breaking parameter $F$ has a dimension of mass squared, $M_{\gamma\gamma}\equiv M_1\cos^2\theta_W+M_2\sin^2\theta_W$ with $M_i$, $i=1,2,3$ being the gaugino masses for all the three SM gauge groups ($U(1)_Y$, $SU(2)_W$ and $SU(3)_c$, respectively) and $v=175$\,GeV is the Higgs field vacuum expectation value. For the soft supersymmetry breaking trilinear couplings entering\,\eqref{Lagr-quarks} we utilize the approximation $A_{Q,l} y_{ij}^{q,l}$ with dimension-of-mass parameters $A_{Q,l}$ defining the overall scale of the trilinear terms in the squark and slepton sectors and dimensionless parameters $y_{ij}^{q,l}$ describing the flavor violation there.  In what follows we consider effect of flavor conserving and flavor violating interactions separately. For the former case we assume the corresponding parameters to be proportional to the fermion masses, $vy_{ii}^{q,l}=m^{q,l}_i$. It is worth to note that we intentionally simplify the structure of sgoldstino couplings to SM fermions, which in general is more complicated, see e.g.\,\cite{Gorbunov:2000cz}.

Note, that coupling constants entering the effective lagrangian\,\eqref{Lagr-photons},\eqref{Lagr-gluons} and~\eqref{Lagr-quarks}  appear via  matching with a microscopic theory at (an unknown) high energy scale. Therefore,  to use the couplings for calculations at low energies one should take into account the renomalization group evolution. Details of this evolution depend not only on the scale at which the effective theory\,\eqref{Lagr-photons},\eqref{Lagr-gluons} and~\eqref{Lagr-quarks} is formulated but also on masses of the SM superpartners. Its precise description is beyond the scope of this study. Still below we include a conservative estimate for leading effects of this scale dependence.  

\subsection{Sgoldstino decay modes}
\label{Subsec:decays}

Couplings of scalar and pseudoscalar sgoldstino to the SM fields induce their decays to the lighter SM particles. Various decay channels involving light sgoldstinos were studied earlier in numerous papers, see e.g.~\cite{Gorbunov:2000th,Gorbunov:2000cz,Gorbunov:2002er,Demidov:2011rd,Astapov:2015otc}. Below we present expressions (some with our corrections) for the relevant decay widths. 

Sgoldstino decays into lepton pairs are governed by\,\eqref{Lagr-quarks}. The corresponding decay widths are 
\begin{equation}\label{decay_leptons}
    \Gamma(S(P) \rightarrow l \, l) = \frac{m_{S(P)} A_l^2 m_l^2}{16 \pi F^2} \times \sqrt{1-\frac{4m_l^2}{m_{S(P)}^2}}\,.
\end{equation}
Decays into a couple of different leptons are typically suppressed. 
Sgoldstinos can decay directly into photons via interactions\,\eqref{Lagr-photons} with the rate 
\begin{equation}\label{decays-to-photons} 
    \Gamma(S(P) \rightarrow \gamma \, \gamma) = 
    R_{S(P)}^{\gamma}\times 
    \frac{m_{S(P)}^3 M_{\gamma \gamma}^2}{32 \pi F^2}\,.
\end{equation}
Here the factor $R_{S(P)}^{\gamma}$ accounts for the scale dependence of coupling in\,\eqref{Lagr-photons}, see e.g.\,\cite{Voloshin:1980zf,Gorbunov:2000th} and has the form 
\begin{equation}\label{rg_factor}
    R_{S}^{\gamma} = \left( \frac{\alpha(M_{\gamma\gamma}) \cdot \beta(\alpha(m_{S(P)}))}{\beta(\alpha(M_{\gamma\gamma})) \cdot \alpha(m_{S(P)})}\right)^2\,,\;\;\;
    R_{P}^{\gamma} = \left( \frac{\alpha(m_{S(P)})}{\alpha(M_{\gamma\gamma}))}\right)^2\,,
\end{equation}
with $\beta(\alpha)$ being electromagnetic $\beta$-function. For pseudoscalar\,\footnote{Also scalar, if parity in sgoldstino sector is violated, see Ref.\,\cite{Gorbunov:2000cz} for  discussion.} sgoldstino there is another contribution to this decay which appears due to sgoldstino mixing with neutral flavourless pseudoscalar mesons via its couplings to gluons\,\eqref{Lagr-gluons} and quarks\,\eqref{Lagr-quarks}. The corresponding partial decay width reads~\cite{Gorbunov:2000th}
\begin{equation}\label{eq-pseudoscalarphotonpi}
\begin{split}
    \Gamma(P \rightarrow \pi^0 \rightarrow \gamma \gamma) = & \left( \frac{\sqrt{2}\pi \epsilon \, M_3 f_{\pi} m_\pi^2}{4\alpha_s(M_3)F} + \frac{B_0 f_\pi}{2F}(m_u - m_d)A_Q \right)^2 \\
    & \times \frac{1}{(m_P^2-m_{\pi^0}^2)^2} \times \Gamma_{tot}(\pi^0) \frac{m_P^3}{m_{\pi^0}^3}\,,
\end{split}    
\end{equation}
\begin{equation}\label{eq-pseudoscalarphotoneta}
\begin{split}
    \Gamma(P \rightarrow \eta \rightarrow \gamma \gamma) = & \left( \frac{\sqrt{2}\pi \, M_3 f_{\pi} m_\eta^2}{4\alpha_s(M_3)F} + \frac{B_0 f_\pi}{2\sqrt{2}F}(m_u + m_d - 2m_s)A_Q \right)^2 \\
    & \times \frac{1}{(m_P^2-m_{\eta^0}^2)^2} \times \Gamma_{\gamma\gamma}(\eta) \frac{m_P^3}{m_{\eta^0}^3}\,,
\end{split}    
\end{equation}
where $f_{\pi}$ is the pion decay constant and $\alpha_s(M_3)$ is the strong coupling constant evaluated at the gluino mass scale, $\Gamma_{tot}(\pi^0)$ stands for the pion total decay rate, $\Gamma_{\gamma\gamma}(\eta)$ stands for the $\eta$-meson decay rate into photons and $\epsilon \equiv \frac{m_u-m_d}{m_u+m_d}$.  
Below for parameter $B_0$ which is related to the quark condensate, we use the value $B_0=\frac{M_{K^0}^2}{m_d+m_s}=5.54$\,GeV and also we take $m_u=2.16$ MeV, $m_d=4.67$\,MeV, $m_s=93$\,MeV for masses of up-, down- and strange quarks respectively\,\cite{Zyla:2020zbs}. 
In general there is an interference between different contributions, see Eqs.~\eqref{eq-pseudoscalarphotonpi}, ~\eqref{eq-pseudoscalarphotoneta}  and ~\eqref{decays-to-photons}, to the decay amplitude of sgoldstino to photons. However, for sets of parameters we use in the study it is negligible, therefore we omit it to simplify the formulas.

Now let us turn to discussion of hadronic decay modes of sgoldstinos. The latter couple directly to both quarks and gluons. For relatively light sgoldstinos one can use the Chiral Perturbation Theory to find the decay widths  for different hadronic final states: meson pairs and trios for the scalar and pseudoscalar sgoldstino assuming parity conservation. 
We derive scalar sgoldstino decay widths to pair of pions using the method advocated in Refs.\,\cite{Shifman1990,Bezrukov_2010}, 
\begin{equation}
\begin{split}
    \Gamma(S \rightarrow \pi^+ \, \pi^-) = 2 \times & \,\Gamma(S \rightarrow \pi^0 \, \pi^0) = 2 \times \frac{1}{32\pi m_S} \times \sqrt{1-\frac{4m_\pi^2}{m_S^2}} \\
    & \times \l \frac{\sqrt{2}(m_S^2+m_\pi^2)}{9F} \l \frac{2 \pi M_3}{\alpha(M_3)} - A_Q \r - \frac{m_\pi^2}{\sqrt{2}} \l \frac{A_Q}{F} + \frac{sin\theta}{v} \r \r^2\,,    
\end{split}
\end{equation}
and to kaons
\begin{equation}
\begin{split}
    \Gamma(S \rightarrow K^+ \, K^-) = & \Gamma(S \rightarrow K^0 \, \bar{K}^0) =  \frac{1}{16\pi m_S} \times \sqrt{1-\frac{4m_K^2}{m_S^2}}  \\
    & \times \l \frac{\sqrt{2}(m_S^2+m_K^2)}{9F} \l \frac{2 \pi M_3}{\alpha(M_3)} - A_Q \r - \frac{m_K^2}{\sqrt{2}} \l \frac{A_Q}{F} + \frac{sin\theta}{v} \r \r^2\,.
\end{split}    
\end{equation}
with the mixing angle between sgoldstino and the Higgs boson\,~\cite{Astapov:2014mea}
\begin{equation}
\label{eq-mixing}
    \theta = -\frac{X}{Fm_h^2}\,,\;\;\;
    X = 2\mu^3 v \sin{2\beta} + \frac{1}{2}v^3(g_1^2M_1 + g_2^2M_2)\cos{2 \beta}\;.
\end{equation}
For $m_h = 125$\,GeV, $\mu = 1$\,TeV, $\tan \beta = 6$, $g_1 = 0.349$, $g_2 = 0.654$, 
$M_1 = 100$\,GeV, $M_2 = 250$\,GeV eq.\,\eqref{eq-mixing} yields for the mixing parameter $X \sim 10^{11}$\,GeV$^4$.

Then, pseudoscalar sgoldstino (if parity is conserved) can't decay into a pair of mesons. Instead, the dominant hadronic decay modes are those into three mesons, which rate can be written as 
\begin{equation}
\begin{split}
    &\Gamma(P \rightarrow 3 \, \text{mesons}) =  \left( \frac{\sqrt{2}\pi \epsilon \, M_3 f_\pi m_\pi^2}{4\alpha_s(M_3)F} + \frac{B_0 f_\pi}{2F}(m_u - m_d)A_Q \right)^{\!2} \!\times \frac{\Gamma(\pi^* \rightarrow 3 \, \text{mesons})}{(m_P^2-m_\pi)^2} \\
    & + \left( \frac{\sqrt{2}\pi \, M_3 f_\pi m_\eta^2}{4\alpha_s(M_3)F} + \frac{B_0 f_\pi}{2\sqrt{2}F}(m_u + m_d - 2m_s)A_Q \right)^{\!2} \!\times \frac{\Gamma(\eta^* \rightarrow 3 \, \text{mesons})}{(m_P^2-m_\eta)^2}\,,\label{eq-pseudoscalarmeson}
\end{split}
\end{equation}
where we introduced decay rates for virtual pion and $\eta$-meson whose mass equals the sgoldstino mass. These rates are governed by the corresponding matrix elements as follows\footnote{We rederived and corrected the expressions from Ref.\,\cite{pich2020effective}.}\,\cite{pich2020effective}
\begin{eqnarray}
     \langle \pi^{0*} | \mathcal{O}_4^{\pi^0} | \pi^0 \pi^0 \pi^0 \rangle & = & \frac{2m_\pi^2}{f_\pi^2}\,, \\
     \langle \pi^{0*} | \mathcal{O}_4^{\pi^0} | \eta \pi^0 \pi^0 \rangle & = & \frac{2m_\pi^2 \epsilon}{\sqrt{3}f_\pi^2}\,, \\
    \langle \pi^{0*} | \mathcal{O}_4^{\pi^0} | \pi^0 \eta \eta \rangle & = & \frac{2m_\pi^2}{3f_\pi^2}\,, \\
    \langle \pi^{0*} | \mathcal{O}_4^{\pi^0} | \pi^0 \pi^+ \pi^- \rangle & = & \frac{2}{3f_\pi^2}(6p_{\pi^+}p_{\pi^-} + 4m_\pi^2 - p_{\pi^{0*}}^2)\,, \\
    \langle \pi^{0*} | \mathcal{O}_4^{\pi^0} | \eta \pi^+ \pi^- \rangle & = & \frac{2m_\pi^2 \epsilon}{3\sqrt{3}f_\pi^2}\,, \\
    \langle \pi^{0*} | \mathcal{O}_4^{\pi^0} | \pi^0 K^+ K^- \rangle & = & \frac{1}{6f_\pi^2}(6p_{K^+}p_{K^-} - p_{\pi^{0*}}^2 - m_\pi^2 + 5m_K^2)\,, \\
    \langle \pi^{0*} | \mathcal{O}_4^{\pi^0} | \pi^0 K^0 \bar{K^0} \rangle & = & \frac{1}{6f_\pi^2}(6p_{K^0}p_{\bar{K^0}} - p_{\pi^{0*}}^2 - m_\pi^2 + 5m_{\bar{K^0}}^2 + m_\pi(1-\epsilon))\,
\end{eqnarray}
and for $\eta^*$
\begin{eqnarray}
     \langle \eta^* | \mathcal{O}_4^{\pi^0} | \pi^0 \pi^0 \pi^0 \rangle & = & \frac{2m_\pi^2 \epsilon}{\sqrt{3}f_\pi^2}\,, \\
     \langle \eta^* | \mathcal{O}_4^{\pi^0} | \eta \pi^0 \pi^0 \rangle & = & \frac{2m_\pi^2}{3f_\pi^2}\,, \\
    \langle \eta^* | \mathcal{O}_4^{\pi^0} | \pi^0 \eta \eta \rangle & = & \frac{2m_\pi^2}{3f_\pi^2}\,, \\
    \langle \eta^* | \mathcal{O}_4^{\pi^0} | \pi^0 \pi^+ \pi^- \rangle & = & \frac{2m_\pi^2 \epsilon}{3\sqrt{3}f_\pi^2}\,, \\
    \langle \eta^* | \mathcal{O}_4^{\pi^0} | \eta \pi^+ \pi^- \rangle & = & \frac{2m_\pi^2}{3f_\pi^2}\,, \\
    \langle \eta^* | \mathcal{O}_4^{\pi^0} | \pi^0 K^+ K^- \rangle & = & \frac{1}{12\sqrt{3}f_\pi^2}(18p_{K^+}p_{K^-} -3 p_{\pi^{0*}}^2 - m_\pi^2 + 10m_K^2)\,, \\
    \langle \eta^* | \mathcal{O}_4^{\pi^0} | \pi^0 K^0 \bar{K^0} \rangle & = & \frac{1}{12\sqrt{3}f_\pi^2}(18p_{K^0}p_{\bar{K^0}} -3 p_{\pi^{0*}}^2 - m_\pi^2 + 10m_K^2)\,.
\end{eqnarray}
We observe, that the hadronic decay rate of sgoldstino, i.e. into three mesons, is governed by the same combination of model parameters, as that into photons (at $M_{\gamma\gamma}=0$). The hadronic decays dominate over photon decay only for sgolsdtino mass $m_S>1.2$\,GeV. While with smaller masses decay width of \eqref{eq-pseudoscalarphotoneta} and \eqref{eq-pseudoscalarphotonpi} is larger than \eqref{eq-pseudoscalarmeson}.

\begin{figure}[!htb]
\center{\includegraphics[width=0.9\textwidth]{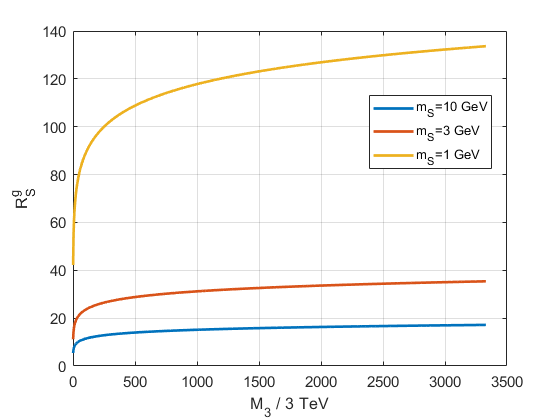}
\caption{Correction factor $R_{S}^{g}\equiv \left( \frac{\alpha_s(M_3) \cdot \beta(\alpha_s(m_{S}))}{\beta(\alpha_s(M_3)) \cdot \alpha_s(m_{S})}\right)^2 $ as a function of gluino mass $M_3$ for a set of sgoldstino masses.}
\label{fig:Renorm}}
\end{figure}

For heavy sgoldstinos it is more appropriate to estimate hadronic decay width using their direct decays into quarks and gluons. Using interaction\,\eqref{Lagr-gluons} one obtains for decays into gluons~\cite{Voloshin:1980zf,Gorbunov:2000th}
\begin{equation}
\label{rate-to-gluons}
    \Gamma(S(P) \rightarrow g \, g) = R_{S(P)}^g\times \frac{m_{S(P)}^3 M_3^2}{4 \pi F^2}\,,
\end{equation}
Here $R^g_{S(P)}$ are the rescaling factors similar to those in~\eqref{rg_factor}
 \begin{equation}\label{rg_factor_gluons}
    R_{S}^{g} = \left( \frac{\alpha_s(M_{3}) \cdot \beta(\alpha_s(m_{S(P)}))}{\beta(\alpha_s(M_{3})) \cdot \alpha_s(m_{S(P)})}\right)^2\,,\;\;\;
    R_{P}^{g} = \left( \frac{\alpha_s(m_{S(P)})}{\alpha_s(M_{3}))}\right)^2\,.
\end{equation}
Note that in the scenario where all the superpartners are not lighter than gluino, (e.g. when all the superpartners are roughly at the same mass scale) only SM particles contribute to the QCD and QED $\beta$-functions in~\eqref{rg_factor} and~\eqref{rg_factor_gluons} and running of couplings $\alpha$ and $\alpha_s$ with renormalization scale, and so the correction $R$-factors in the decay rates~\eqref{decays-to-photons} and~\eqref{rate-to-gluons} are conservatively calculated within the SM. The most important correction here comes from the scale dependence of the strong coupling constant; for a set of scalar sgoldstino masses the corresponding factor is outlined in Fig.\,\ref{fig:Renorm}.   
Sgoldstino decay width to a pair of light quarks is given by the same expression as for decay into leptons~\eqref{decay_leptons} apart from the mass (lepton mass is replaced with the corresponding quark mass) and a factor accounting for the number of colors. Due to suppression by the mass of light quarks this channel is subdominant as compared to decays into gluons and we neglect it in what follows.

It is worth stressing that the formulas above for the meson decay rates of scalar sgoldstino are obtained to the leading order in Chiral Perturbation Theory. It is well known that such estimates for any SM singlet scalar coupled to the gluonic tensor suffer from uncertainties due to the meson strong interactions in the final state, see e.g.\,\cite{Donoghue:1990xh,Bezrukov_2010,Bezrukov:2018yvd}. It has been argued that for the scalar masses in 0.5-1.5\,GeV range the corrections to the hadronic decay rates vary from several to about a hundred \cite{Donoghue:1990xh}. The situation is far from being settled \cite{Bezrukov:2018yvd}. And though one may expect that multihadronic modes are irrelevant in this mass range, the two-meson modes are not under control, especially in the regions close to the hadronic scalar resonances. This whole issue is far beyond the scope of this paper. We simply use the perturbative estimates above for the hadronic modes, keeping in mind that once the issue is resolved we will be able to properly rescale the number of sgoldstino signal events in the FASER detector.

The sgoldstino decay pattern for each sgoldstino  mass is determined by the hierarchy of the soft supersymmetry breaking parameters which enter sgoldstino decay widths. In case of equal soft terms, $M_3=M_{\gamma\gamma}=A_0$, the decays of scalar sgoldstino into hadrons dominate, if kinematically allowed,   while decays into photons do, if not. For pseudoscalar sgoldstinos the hadronic modes are subdominant, while $m_P\lesssim 1$\,GeV. 

\begin{figure}[!htb]  
\centerline{
\includegraphics[width=0.5\linewidth]{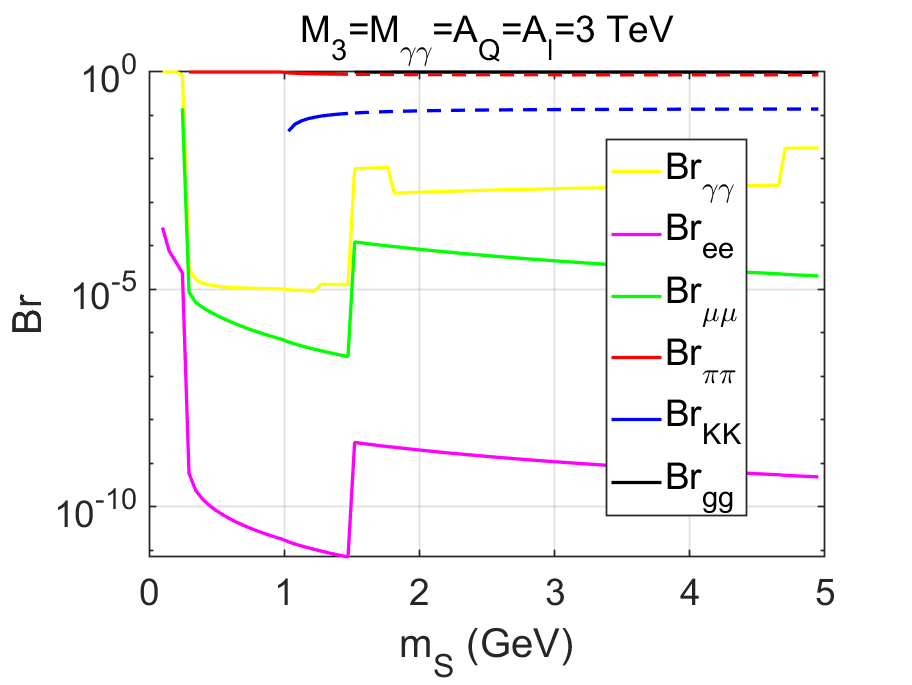}
\includegraphics[width=0.5\linewidth]{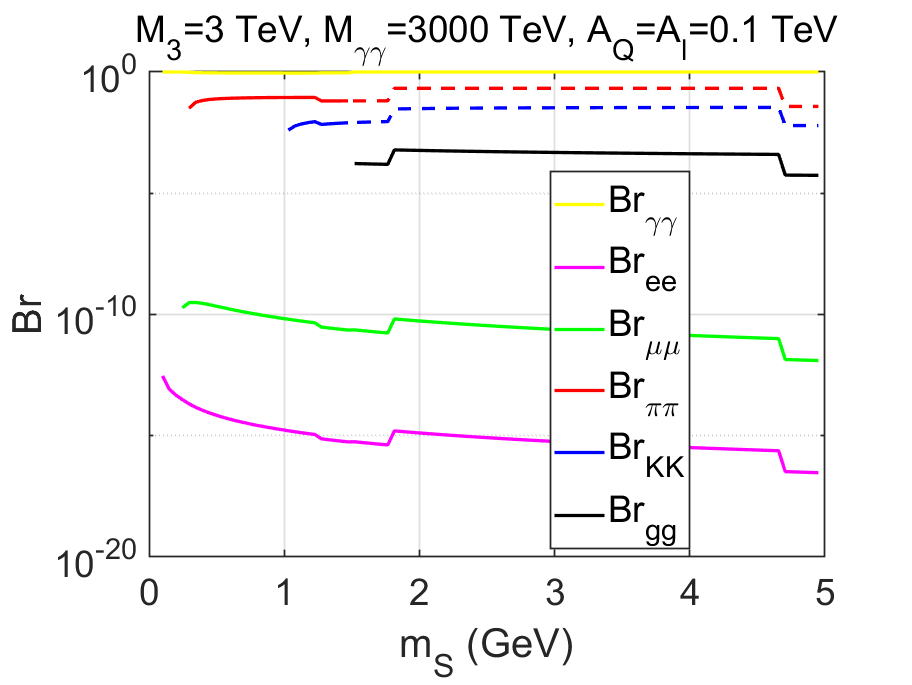}}

\vskip 0.2cm
\centerline{\includegraphics[width=0.5\linewidth]{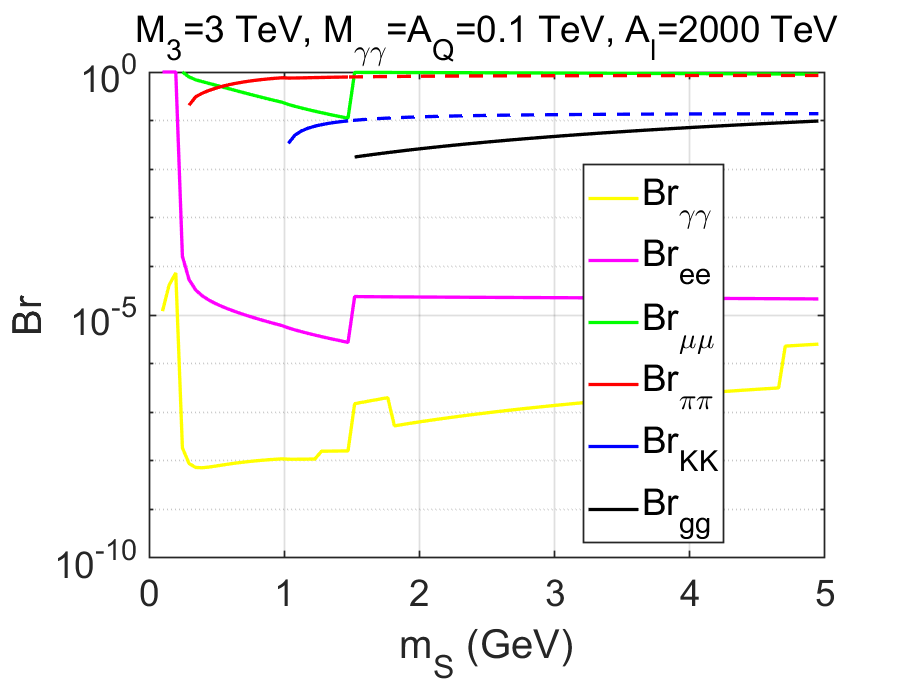}
\includegraphics[width=0.5\linewidth]{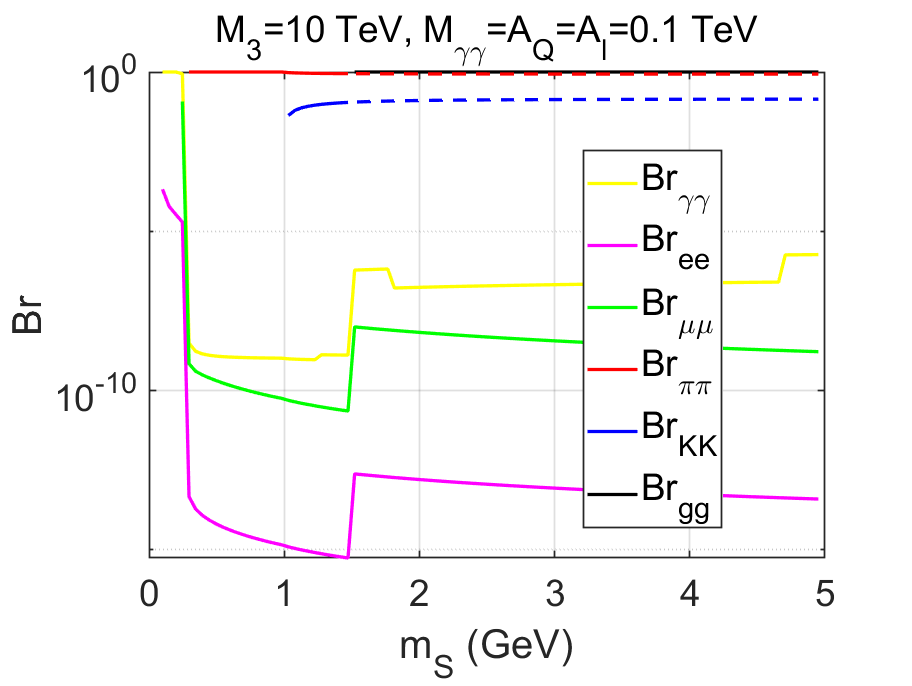}}  
\caption{Scalar sgoldstino decay branchings for channels $S\to ee, \mu\mu, \gamma\gamma, \pi\pi, KK$ for chosen sets of supersymmetry breaking soft terms. For $m_S>1.5$~GeV we use decays into gluons for all hadronic decay modes.}
\label{fig:branchings}
\end{figure}

A particular decay mode may dominate in case of a suitable hierarchical structure of soft parameters. Say, for $m_S=1\;$GeV the decay into photons dominates,
\begin{equation}\label{eq19}
    \text{Br}(S \rightarrow \gamma \gamma) \approx 1\;, \;\;\; \text{if} \;\;\; \frac{M_{\gamma \gamma}}{M_3} > 1000\;, \;\;\frac{M_{\gamma \gamma}}{A_Q} > 0.5\;, \;\;\frac{M_{\gamma \gamma}}{A_l} > 1\;, 
\end{equation}
the decays into mesons dominate, 
\begin{equation}\label{eq20}
    \text{Br}(S \rightarrow mesons) \approx 1\;, \;\;\; \text{when} \;\;\; \frac{M_3}{M_{\gamma \gamma}} > 0.01\;,\;\; \frac{M_3}{A_l} > 0.0025\;, 
\end{equation}
and the decay into muons dominates, 
\begin{equation}\label{eq21}
    \text{Br}(S \rightarrow \mu \mu) \approx 1\;, \;\;\; \text{for} \;\;\; \frac{A_l}{M_3} > 4000\;, \;\;\frac{A_l}{M_{\gamma \gamma}} > 10\;, \;\;\frac{A_l}{A_Q} > 2\;. 
\end{equation}
Lighter sgoldstinos, $m_S<2m_\mu$, mostly decay into photons, e.g. for $m_S=100 \;$MeV
\begin{equation}\label{eq22}
    \text{Br}(S \rightarrow \gamma \gamma) \approx 1\;, \;\;\; \text{if} \;\;\; \frac{M_{\gamma \gamma}}{A_l} > 0.05\;.
\end{equation}
As to the hadronic modes, we treat sgoldstinos as decaying into gluons for $m_{S(P)}>1.5$\,GeV, and decaying into mesons otherwise. 
Scalar sgoldstino decay branchings for four particular sets of soft parameters are presented in Fig.\,\ref{fig:branchings}. 
In what follows we use these four sets as benchmark points for numerical calculations and presentation of our results. 
\subsection{Meson decays into sgoldstino}
\label{Subsec:mesonsig}

Produced in proton collisions mesons can decay into sgoldstino, if it is kinematically allowed. There is also a certain difference between the cases of scalar and pseudoscalar sgoldstino, if their interactions with quarks respect parity (recall, this choice we adopt in this study). Many relevant production modes were investigated in Refs.\,\cite{Gorbunov:2000th} and \cite{Astapov:2015otc}. 

\paragraph{Flavor-conserving couplings.}
We start with meson decay modes into sgoldstino, originated from sgoldstino couplings to SM quarks which conserve the quark flavor. Naturally, these modes should either include the flavorless initial meson, or exploit the weak flavor violating interactions (i.e. charged currents). 

\begin{figure}[!htb]\centerline{
    \includegraphics[width=\textwidth]{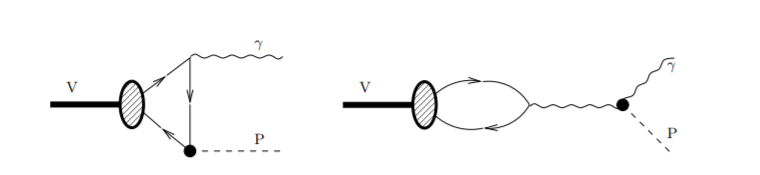}}
    \caption{Feynman diagrams of sgoldstino $P$ production in flavorless meson $V$ decays emanated by flavor-conserving pseudoscalar sgoldstino $P$ couplings \cite{Gorbunov:2000th}. Solid lines indicate quarks forming the meson. Similar diagrams contribute to scalar $S$ production.} 
    \label{fig:meson-flavor-conserving-decays}
\end{figure}

The flavorless vector mesons, $\rho^0$, $\phi$, $\omega$, $J/\psi$, $\Upsilon$, etc, can decay to sgoldstino and photon via diagrams depicted in Fig.\,\ref{fig:meson-flavor-conserving-decays}. 
Following Ref.\,\cite{Gorbunov:2000th}, the corresponding meson branchings can be evaluated from the ratio of these rates to the rates into lepton pairs as 
\begin{equation}
    \frac{\Gamma(V \rightarrow P(S) \gamma)}{\Gamma(V \rightarrow \gamma \rightarrow e^+ e^-)} = \frac{M_V^2(A_Q \pm M_{\gamma \gamma} \, R^\gamma_{P(S)})^2}{16 \pi \alpha F^2}\,,
\end{equation}
where '$-(+)$' stands for decays into (pseudo)scalar and $\alpha$ is the fine-structure constant.   

\begin{figure}[!htb]  
\centerline{\includegraphics[width=0.5\linewidth]{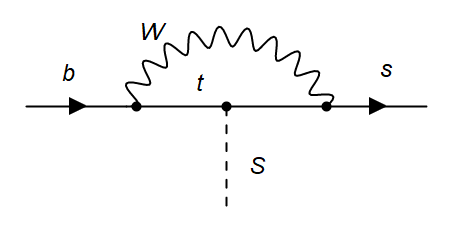} 
\includegraphics[width=0.5\linewidth]{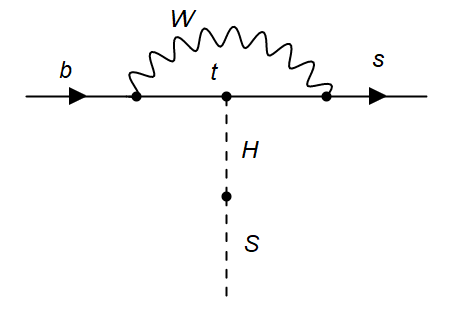} }
\caption{Feynman diagrams for meson decays into sgoldstino via flavor-conserving sgoldstino couplings\,\cite{Gorbunov:2000th}: a) diagram of sgoldstino production via Yukawa-type coupling to top-quark; b) diagram with sgoldstino-higgs mixing.} 
\label{fig:feynman higgs}
\end{figure}

Decays into sgoldstino and mesons of different quark flavors proceed starting at one-loop level with virtual $W$-boson involved, which is responsible for the change of flavor. Sgoldstinos  emerge via coupling to virtual quarks running in the loop or (for scalar sgoldstino only if parity in sgoldstino-couplings are conserved) via mixing with the Higgs boson which couples to the same quarks, see Refs.\,\cite{Astapov:2015otc,Bezrukov_2010}.  In both cases the largest contribution is expected from virtual top-quarks. Examples of the Feynman diagrams for such processes involving $b$ to $s$ transition are presented in Fig.\,\ref{fig:feynman higgs}.
For the beauty meson decays into strange meson and sgoldstino the branching reads\,\cite{Astapov:2015otc}
\begin{equation}\label{eq_BMesonDecay}
    \text{Br}(B \rightarrow X_sS)=0.3 \times \left( \frac{m_t}{m_W} \right)^2 \times \left( 1 - \frac{m_S^2}{m_b^2} \right)^2 \times \left( \frac{A_Qv}{F} + \theta \right)^2
\end{equation}
For this model the sgoldstino-Higgs mixing dominates over the sgoldstino-quark contribution at $A_Q \lesssim 10$\,TeV. 

In the same way, charmed and strange mesons can also decay into sgoldstino. The branchings of the kaons are as follows \begin{eqnarray}
    \text{Br}(K^+ \rightarrow \pi^+ S) & = & 1.3 \cdot 10^{-3} \times \frac{2|\Bar{p}_S|}{m_K} \times \left( \frac{A_Qv}{F} + \theta \right)^2,\\
    \text{Br}(K_L \rightarrow \pi^0 S) & = & 5.5 \cdot 10^{-3} \times \frac{2|\Bar{p}_S|}{m_K} \times \left( \frac{A_Qv}{F} + \theta \right)^2,
\end{eqnarray}
where sgoldstino 3-momentum equals 
\[
|\Bar{p}_S|=\frac{1}{2M_K}\sqrt{\left( \left( M_K-m_S\right)^2-M_\pi^2\right)\left( \left( M_K+m_S\right)^2-M_\pi^2\right)}\,.
\]
Similar branchings of $K_S$ are much smaller\,\cite{Bezrukov_2010}. The charmed meson branchings are suppressed much more due to combinations of the quark mixing parameters and smallness of the $b$-quark Yukawa coupling. The largest contribution for the charmed mesons comes from the tree-level process (in which sgoldstino is emitted from the initial quark) with branching\,\cite{Bezrukov_2010} 
\begin{equation}
    \text{Br}(D \rightarrow e \nu \, S) = 5.7 \cdot 10^{-9} \times \left( \frac{A_Qv}{F} + \theta \right)^2 \times f \left( \frac{m_S^2}{m_D^2} \right)\,,
\end{equation}
where
\begin{equation*}
    f(z) = (1-8z+z^2)(1-z^2)-12z^2\log{z}\,.
\end{equation*}

Remarkably, the diagrams similar to those in Fig.\,\ref{fig:feynman higgs} give rise to the decay $\eta\rightarrow \pi^0 S$ with branching  
\begin{equation}
    \text{Br}(\eta \rightarrow \pi^0 S) = 10^{-6} \times \frac{2|\Bar{p}_S|}{m_\eta} \times \left( \frac{A_Qv}{F} + \theta \right)^2.
\end{equation}

There is also process $\eta \rightarrow \gamma \gamma P$ described in \cite{2018ALPFASER}, but the contribution coming from this decay is negligibly small.

For the pseudoscalar sgoldstino the effective lagrangian describing $b\to s$ transition looks (neglecting contributions suppressed as $m_s/m_b$, see e.g.~\cite{Dolan:2014ska,Dobrich:2018jyi})
\begin{equation}
    {\cal L} = h_{sb}^R P\bar{s}_Lb_R+{\rm h.c.}\,
\end{equation}
where
\begin{equation}
    h_{sb}^R=\frac{\alpha m_b m^2_t A}{4\pi m_W^2\sin^2{\theta_W}F}V_{tb}V^*_{ts}\log{\frac{\Lambda^2}{m_t^2}}\,,
\end{equation}
and $\Lambda$ is a relevant new physics scale which is in our case should be taken of order $\sqrt{F}$. The  partial widths for the relevant flavour violating decays of pseudoscalar $B$-mesons read
\begin{equation}
    \Gamma\left(B\to K P\right) = \frac{|h_{sb}^R|^2}{64\pi m_B^3}\lambda(m_B^2, m_K^2, m_P^2)\left|f^{B_0}_0(m_P^2)\right|^2 \l \frac{m_B^2-m_K^2}{m_b-m_s} \r^2\,,
\end{equation}
\begin{equation}
    \Gamma\left(B\to K^* P\right) = \frac{|h_{sb}^R|^2}{64\pi m_B^3}\lambda^{3}(m_B^2, m_K^2, m_P^2)\left|A^{B_0}_0(m_P^2)\right|^2\frac{1}{(m_b+m_s)^2}\,,
\end{equation}
where 
\begin{eqnarray}
    \lambda(x,y,z) \equiv \sqrt{(x-y-z)^2 - 4yz}\,, \\
    f^{B_0}_0(q^2) = \frac{0.33}{1-\frac{q^2}{38\, \text{GeV}^2}}\,, \\
    A^{B_0}_0(q^2) = \frac{1.36}{1-\frac{q^2}{28 \, \text{GeV}^2}} - \frac{0.99}{1-\frac{q^2}{37 \,. \text{GeV}^2}}\,.
\end{eqnarray}

\paragraph{Flavor-violating couplings.}
We proceed by considering the sgoldstino production channels initiated by flavor-violating sgoldstino interactions to quarks. Corresponding coupling constants are proportional to off-diagonal entries of the left-right squark mass squared matrix, and we approximate it as
\[
A_Q v y^q_{ij}\equiv m^{2\;LR}_{ij}\,.
\]
Sgoldstino is neutral, and hence the mixing involves either two up-type quarks or two down-type quarks. 

The most promising sgoldstino sources are relatively long-lived mesons. A pseudoscalar meson $M$ decays into a lighter one $X$ and scalar sgoldstino with rate 
\begin{equation}\label{eq:meson_decay}
    \Gamma(M \rightarrow XS) = F_{M \rightarrow X}^2 \left( \frac{m_M^2-m_X^2}{m_{q_i}-m_{q_j}} \right)^2 \frac{\lambda(m_M^2, m_X^2, m_S^2)}{32 \pi^2 m_M^3} \frac{m_{q \, ij}^{LR \, 4}}{F^2}\,.
\end{equation}
Here $m_{q_i}$ are the quark masses and 
$F_{M \rightarrow X}$ are dimensionless form factors. The latter can be calculated for various meson transitions using formulas from Ref.\,\cite{Palmer_2014}:
\begin{equation}
    F_{M \rightarrow X}(0) = \frac{F_{D \rightarrow X}(0) \times m_D^{3/2}}{m_M^{3/2}}\,,
\end{equation}
\begin{equation}\label{eq67}
    F_{M \rightarrow X}(m_S^2) = \frac{F_{M \rightarrow X}(0)}{\left( 1 - \frac{m_S^2}{m_{M^*}^2} \right)\left( 1 - \frac{\tilde{\alpha} m_S^2}{m_{M^*}^2} \right)}\,.
\end{equation}
The expression~\eqref{eq67} includes a pole at the mass of the first excited state $m_{M^*}$ and another one at $\tilde{\alpha}\, m_{M^*}$, where the dimensionless factor $\tilde{\alpha}$ accounts for contributions of higher resonances. 
Using~\eqref{eq:meson_decay} one can obtain partial decay widths for $B\to KS$ and $D_s\to KS$ transitions as 
\begin{gather}
    \Gamma(D_s \rightarrow KS) = F_{D_s \rightarrow K}^2 \left( \frac{m_{D_s}^2-m_K^2}{m_c-m_u} \right)^2 \frac{\lambda(m^2_{D_s},m_K^2,m_S^2)}{32 \pi^2 m_{D_s}^3} \frac{m_{U \, 12}^{LR \, 4}}{F^2}\,, \\
    \Gamma(K_s \rightarrow \pi S) = F_{K_s \rightarrow \pi}^2 \left( \frac{m_{K_s}^2-m_\pi^2}{m_s-m_d} \right)^2 \frac{\lambda(m^2_{K_s},m_\pi^2,m_S^2)}{16 \pi^2 m_{K_s}^3} \frac{(m_{D \, 21}^{LR \, 2} + m_{D \, 12}^{LR \, 2})^2}{F^2}\,, \\
    \Gamma(\eta' \rightarrow K S) = F_{\eta' \rightarrow K}^2 \left( \frac{m_{\eta'}^2-m_K^2}{m_s-m_d} \right)^2 \frac{\lambda(m^2_{\eta'},m_K^2,m_S^2)}{48 \pi^2 m_{\eta'}^3} \frac{m_{D \, 12}^{LR \, 4}}{F^2}\,.     
\end{gather}

\paragraph{Direct limits on meson decays into sgoldstino.}

We finish the study of sgoldstino production modes by noting that many of the modes above are constrained from thorough investigations of the meson decays, which typically include searches for new light particles.  A summary of the relevant upper limits are presented in Tab.\,\ref{tab:tab4}. 
\begin{table}[htb!]
    \begin{center}
    \label{tab:tab4}
    \begin{tabular}{| l | l | l | l |}
    \hline
      & $S \rightarrow \mu^+ \mu^-$ & $S \rightarrow \gamma \gamma$ & $S \rightarrow mesons(2\pi^0, \pi^+ \pi^-, K^+K^-)$ \\
      \hline
    Br$(B \rightarrow X_s S)$ & $2 \cdot 10^{-10}$ & -- & $10^{-6}$\\ 
    Br$(D \rightarrow X_s S)$ & $4.3 \cdot 10^{-6}$ & -- & $10^{-6}$ \\
    Br$(\eta \rightarrow \pi S)$ & $5 \cdot 10^{-6}$ & $3 \cdot 10^{-5}$ & $3 \cdot 10^{-3}$\\
    Br$(K \rightarrow \pi S)$ & $1.5 \cdot 10^{-9}$ & $1.3 \cdot 10^{-8}$ & $10^{-7}$\\
    \hline
    \end{tabular}
    \end{center}
\caption{Upper limits on meson branchings into sgoldstino  \cite{Zyla:2020zbs,Gorbunov:2000th,PhysRevD.79.092004,Wei_2009}.}
\end{table}
In our investigation we always consider the regions in the model parameter space which are consistent with bounds of Tab.\,\ref{tab:tab4}. Typically, for all other model parameters fixed, these bounds impose upper limits on the off-diagonal entries in the sfermion squared mass matrices. 

\section{Light sgoldstino flux at FASER}
\label{Sec:flux}

\subsection{FASER in brief}
\label{Sec:FASER}

ForwArd Search ExpeRiment (FASER) at the LHC has been proposed\,\cite{Feng_2018} and then  rapidly developed\,\cite{Ariga_2019,fasercollaboration2019faser} and approved to be installed in the LHC tunnel at a distance of $L_{min}=480$\,m from the proton collision point of the ATLAS detector. Current FASER schedule contains two operational stages, FASER-I and FASER-II, coordinated in time with LHC run 3 (LHC-3) and High Luminosity LHC (HL-LHC). Apart from different statistics of the proton collisions (the expected integrated luminosity is $L=150$\,fb$^{-1}$ and $L=3000$\,fb$^{-1}$ respectively), the experiment at different stages will be equipped with different detectors. The latter are both of the cylindrical form with axis on the straight line crossing the proton collision point, with radii of $R=0.1$\,m and 1\,m and lengths of $\Delta L\equiv L_{max}-L_{min}=1.5$\,m and 5\,m for FASER-I and FASER-II, respectively. From the proton collision point, the detector will be visible at an angle of $\Theta=2R/L_{min}\approx 4.2\times 10^{-4} (4.2\times 10^{-3}$) for FASER-I(II).  The detector of FASER-II has much larger fiducial volume than FASER-I and with twenty times higher integrated luminosity generically will have much higher sensitivity to new physics, which we demonstrate in due course on our example of supersymmetric model with light sgoldstinos. The relevant for our study geometrical parameters are summarized in Tab.\,\ref{FASERPARAM}. { It is worth mentioning that within Future Forward Facility program\,\cite{Feng:2022inv} both the position and geometrical size of FASER-II are under debates. We choose here the numbers from the original proposal and its subsequent development, but with new ideas implemented in the FASER project the presented in our paper estimates must be refined.}
\begin{table}[!htb]
    \begin{center}
    \label{FASERPARAM}
    \begin{tabular}{| l | l | l |}
    \hline
     & FASER1 & FASER2 \\
     \hline
    $L_{min}$ & $480 \, \text{m}$ & $480 \, \text{m}$ \\
    $L_{max}$ & $481.5 \, \text{m}$ & $485 \, \text{m}$ \\
    $R$ & $0.1 \, \text{m}$ & $1 \, \text{m}$ \\
    $\Theta$ & $0.00042$  & $0.0042$ \\
    $L$ & $150 \, \text{fb}^{-1}$ & $3000 \, \text{fb}^{-1}$ \\
    \hline
    \end{tabular}
    \end{center}
 \caption{Geometrical parameters of FASER-I and FASER-II 
    \cite{fasercollaboration2019faser}.}
    \end{table}

The detector systems of the experiment are thoroughly described in Ref.\,\cite{fasercollaboration2019faser}. FASER is designed to recognize photons and charged particles. The former are converted inside the lead layers  producing the electromagnetic shower which can be measured, while the latter can be recognized with the help of deflecting magnets, the tracking system and the downstream electromagnetic calorimeter. The veto system includes upstream scintillators and gets rid of entering charged particles.    

Sgoldstinos can be observed in FASER detector if they decay inside into a pair of charged particles or photons. The probability of a relativistic particle with large $\gamma$-factor and lifetime $\tau$ to reach the distance $L$ is $P(L)\equiv \text{e}^{-L/(\tau\gamma)}$. Sgoldstino must reach the FASER detector and decay inside it. Hence, for the contributions to sgoldstino fluxes associated with its direct production and production in decays of short-lived mesons (one can call the both cases as a prompt production), the probability of sgoldstino to decay inside the FASER detector is 
\[
P_p= \text{e}^{-L_{min}/(\tau\gamma)}- \text{e}^{-L_{max}/(\tau\gamma)}\;,
\]
while for the contribution coming from decays of long-lived mesons happened at a distance $l$ from the proton collision point (non-prompt production) the probability is 
\[
P_{np}= \text{e}^{-(L_{min}-l)/(\tau\gamma)}- \text{e}^{-(L_{max}-l)/(\tau\gamma)}=P_p\times \text{e}^{l/(\tau\gamma)}\,.
\]
For the interesting case of long-lived sgoldstinos, having enough time to reach the FASER detector and decay length definitely exceeding the length of FASER detector $L_{max}$-$L_{min}$, these formulas get simplified as  
\[
P_p=\text{e}^{-L_{min}/(\tau\gamma)} \times \frac{L_{max}-L_{min}}{\tau\gamma} \,,\;\;\;\;\;
P_{np}=\text{e}^{-(L_{min}-l)/(\tau\gamma)} \times \frac{L_{max}-L_{min}}{\tau\gamma}\;, 
\]
and for sufficiently long-lived sgoldstinos the exponents approach unity. The strong suppression of sgoldstino decay inside the detector volume can be avoided with both factors---exponential and linear in sgoldstino width $1/\tau$ -- not very far from unity. The optimal case then would be for sgoldstino with decay length of the order of the distance to the FASER detector, which is about $500$\,m. That implies $1/\tau\simeq 10^{-15}-10^{-14}$\,GeV for typical $\gamma=30-300$.

There are three main contributions to the 
sgoldstino flux at FASER, which should be investigated separately. The first contribution is from direct sgoldstino production in proton-proton collisions, which is determined mostly by sgoldstino couplings to gluons~\eqref{Lagr-gluons}. Sgoldstinos appear at the collision point, and we need to know their distribution over the transverse and longitudinal  momenta to understand which trajectory crosses the FASER detector. 
The second contribution is from decays of the short-lived mesons produced in proton collisions. Sgoldstinos are secondary particles here, and the momentum distribution of their parent mesons impacts on their trajectories and chances  to cross the FASER detector. Finally, the third contribution comes from decays of long-lived mesons like kaons, which are abundantly produced in proton collisions, but cover a macroscopic distance before decays. In this case sgoldstinos are produced at some distance from the proton collision point, which also should be accounted when estimating the sgoldstino chances to reach the FASER detector. All the three contributions are considered below in detail. We also found that high energy photons originated from proton collisions can propagate along the beamline and then produce sgoldstinos in scattering off media through the Primakoff effect. The process is interesting only in models with dominant sgoldstino coupling to photons, and we discuss it in the proper place.

\subsection{Direct sgoldstino production in proton collisions}
\label{Subsec:direct}

\begin{figure}[!htb]
    \centerline{\includegraphics[width=0.48\textwidth]{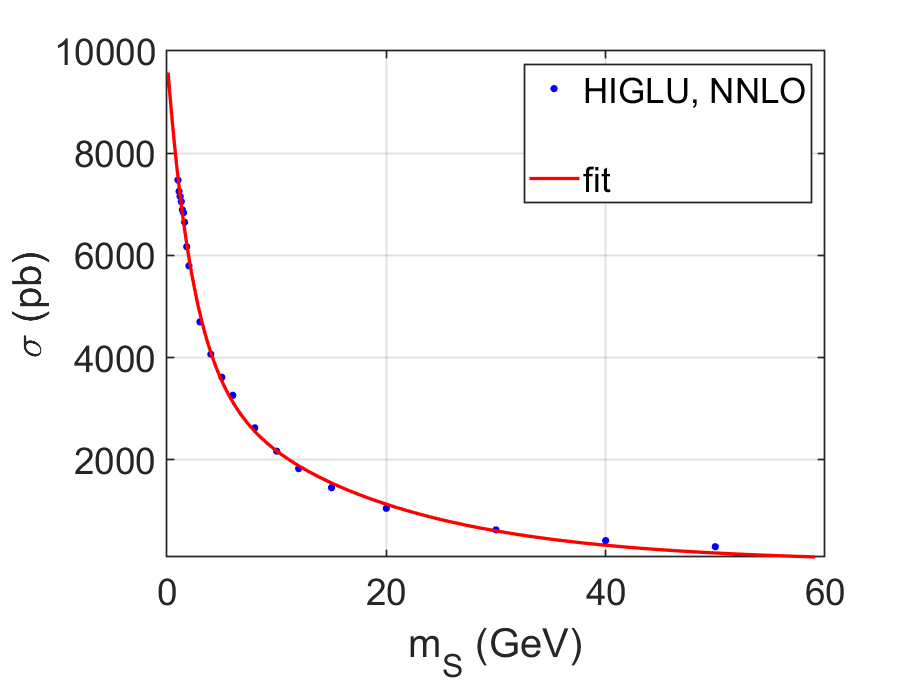}\hskip 0.04\textwidth \includegraphics[width=0.48\textwidth]{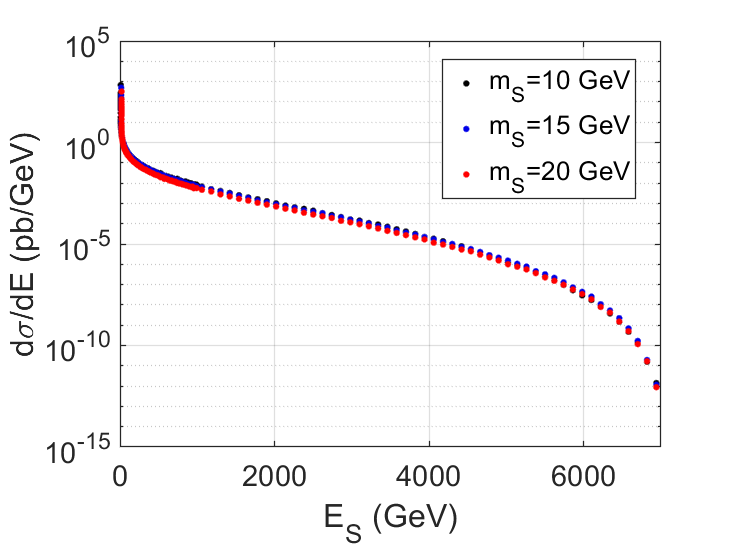}}
    \caption{{\it Left panel:} Sgoldstino production cross section in the gluon fusion at $\sqrt{s}=14$\,TeV is evaluated for $M_3=3$\,TeV and $\sqrt{F}=10$\,TeV. {\it Right panel:} Differential cross section as a function of sgoldstino energy $E_S$ for a set of sgoldstino masses $m_S$.}
    \label{fig:direct}
\end{figure}

Sgoldstinos can be produced in proton collisions directly via the gluon fusion mechanism which involves the sgoldstino coupling to gluons~\eqref{Lagr-gluons}. This process resembles the SM Higgs boson production in the limit of the point-like interactions to gluons and here we use the HIGLU package~\cite{HIGLU} to calculate sgoldstino production cross section in proton-proton collisions at $\sqrt{s}=14$~TeV at NNLO in QCD using NNPDF3.1~\cite{NNPDF:2017mvq} for the parton distribution functions and by properly rescaling the effective Higgs boson coupling to gluons and varying its mass. This allows one to evaluate the sgoldstino production rate at the renormalization scale as low as $Q=1$\,GeV. We put this scale equal to the sgoldstino mass $m_S$ and obtain the total production cross section of sgoldstino at LHC with proton beam energy $\sqrt{s}=14$\,TeV and $M_3=3$\,TeV, $\sqrt{F}=10$\,TeV,  
as depicted on the left panel of Fig.\,\ref{fig:direct}. 
The estimates for other model parameters are obtained straightforwardly by the rescaling of sgoldstino coupling to gluons\,\eqref{Lagr-gluons}, which is proportional to $\alpha_s(M_3)/\beta(\alpha_s(M_3))\times M_3/F$, see eqs.\,\eqref{rate-to-gluons}, \eqref{rg_factor_gluons}. Therefore, for the sgoldstino production cross section one finds
\[
\sigma\l M_3,\sqrt{F}\r = \l \frac{\alpha_s(M_3) \cdot \beta(\alpha_s(3 \, \text{TeV}))}{\beta(\alpha_s(M_3)) \cdot \alpha_s(3 \, \text{TeV})} \r^2 
     \l \frac{10 \, \text{TeV}}{\sqrt{F}} \r^4 \l \frac{M_3}{3 \,\text{TeV}} \r^2
    \times \sigma\l 3\,\text{TeV},\, 10\,\text{TeV} \r
\]
and the last factor stands precisely for the cross section shown on the left panel of Fig.\,\ref{fig:direct}. 

The energy spectra of the produced sgoldstinos are estimated by fitting to the spectra obtained by making use of the CompHEP package\,\cite{Pukhov:1999gg,Gorbunov:2001pd}. They are presented on the right panel of Fig.\,\ref{fig:direct}. We found a reasonably accurate numerical fit to the sgoldstino distribution over 3-momentum $p$ { which reads} 
\begin{equation}
    \sigma^{-1} \times \frac{d\sigma}{dp} = \frac{0.21}{\text{GeV}} \times \left( 1 - 4.76\times 10^{-4}\times\left( \frac{p^2+m_S^2}{\text{GeV}^2} \right)^{0.15487} \right)^{2.204\times 10^3} 
    \label{fitdsdp}
\end{equation}
{To estimate } the production cross section of sgoldstinos lighter than 1\,GeV we use the numerical extrapolation of the total cross section from the masses above 1\,GeV using the following fitting formula 
\begin{equation}
    \sigma(m_S) = 6\,\text{nb}\times e^{-0.43\,\frac{m_S}{\text{GeV}}} + 3.9\,\text{nb}\times e^{-0.062\,\frac{m_S}{\text{GeV}}}\,.
\end{equation}
The result is shown by the red curve on the left plot of Fig.\,\ref{fig:direct}.  We use it as our estimate of total scalar and pseudoscalar sgoldstino production cross section in this study.

Finally, for the sgoldstino distribution over transverse momenta $p_T$ we adopt the normalized to unity distribution of mesons obtained with EPOS generator of hadronic interactions\,\cite{Pierog_2015,internetCRMC}, i.e.  
\begin{equation}
    f_S(p_T)dp_T = \frac{\bar\alpha^2\,p_Tdp_T}{\Lambda^2+m_S^2} \times e^{-\frac{\bar\alpha p_T}{\sqrt{\Lambda^2+m_S^2}}}\,, 
\label{pT}
\end{equation}
where $\Lambda = 100$\,MeV for the QCD scale and $\bar\alpha=1.3$ in agreement with simulation data of Cosmic Ray Monte Carlo (CRMC) package\,\cite{Pierog_2015,Aaij_2015,internetCRMC}. 

To be detected the sgoldstino must pass through the detector, so its trajectory must deviate from the beam line by an angle less than the detector viewing angle, see Tab.\,\ref{tab:tab4}, that implies the following bound on the sgoldstino momentum  
\begin{equation}
\label{eq:angle-limit}
\frac{p_T}{p}<\frac{\Theta}{2}\;.
\end{equation}
All sgoldstinos obeying this bound contribute to the sgoldstino flux at FASER.  Finally, the number of sgoldstino events in the detector for the LHC collecting the integral luminosity~$L$ in case of direct sgoldstino production reads
\begin{equation}\label{eq:direct_prod}
 N^{\rm direct}_S= L\times \int_{0}^{\sqrt{s}} dp \frac{d\sigma(pp\to S)}{dp} \int_0^{\frac{1}{2}\Theta p}   f_S(p_T)dp_T \times P_p(p)\,.
\end{equation}

{Let us note that in the above study we neglect parton sub-processes with gluon and quark radiation in the final state. One can expect that they give sub-leading contribution to the production of sgoldstino in forward direction, see e.g.~\cite{Bakhti:2020vfq}. At the same time,  our calculation involves application of parton distribution functions at very low $Q^2$ and small $x$. This leads to large and sometimes poorly controllable uncertainties in the cross section predictions especially for very light sgoldstinos. Performing variation of the factorization scale (we take it equal to the renormalization scale) we estimate the size of possible corrections. They are highly asymmetric. Namely, {\it increasing} the scale from sgoldstino mass $M_S$ to $2M_S$ we observe from tens to hundred percent corrections even for very light sgoldstinos, $M_S=1.5$-$1$\,GeV, respectively. However, the corrections become much larger, factors 2-200 for {\it decreasing} factorization scale from $M_S$ to $M_S/2$ and the same sgoldstino mass interval. The direct production cross section for lighter sgoldstinos is obtained via extrapolation as discussed above and hence even larger uncertainties are expected in this case. Similar to the hadronic decay widths of very light sgoldstinos (see discussion in Section~2.2) further study is required to obtain reliable predictions in this case. One possible way to improve the situation here may be found by applying the $k_T$-factorization approach with BFKL equations to sum up $\log{1/x}$ contributions in PDFs, see~\cite{Feng:2022inv} for real examples from the FASER physics program.}

{An impact of these uncertainties is different for different parameters of the model. For the considered setup, the number of signal events in the detector scales as $\frac{M_{soft,1}^2M_{soft,2}^2}{F^4}$, where $M_{soft,1}$ and $M_{soft,2}$ are soft SUSY breaking parameters entering the sgoldstino production cross section and the decay rate, respectively. Therefore, for instance, the theoretical uncertainty in production cross section of two orders of magnitude translates to the uncertainty within a factor of few in the sensitivity plots on $(m_{S(P)},\sqrt{F})$ plane. 
}

\subsection{Sgoldstino production via Primakoff process}
\label{Subsec:Primakoff}
Another promising source of { sgoldstino} at FASER is inelastic photon nuclei scattering, the  Primakoff process. { Its} role for axion-like particle production in FASER has been thoroughly described in \cite{2018ALPFASER}. Using the same method and replacing the coupling constants we get the formula for differential cross section. 
\begin{equation}
\label{eq:Primakoff}
\frac{d\sigma_{Pr}}{d\theta_{P\gamma}} = \frac{M_{\gamma\gamma}^2}{2F^2} \alpha Z_{nuc}^2 F(t)^2 \frac{p_P^2 sin^3(\theta_{P\gamma})}{t^2} 
\end{equation}
with $t = -(p_P - p_{\gamma})^2$ and
\begin{equation}
\label{eq:PrimakoffForm}
F(t) = 
\begin{cases}
   \frac{a^2t}{1+a^2t} & t < 7.39m_e^2 \\
   \frac{1}{1+t/d} & t > 7.39m_e^2
 \end{cases}
\end{equation}
where $a = 111 Z_{nuc}^{-1/3}/m_e$, $d = 0.164 A_{nuc}^{-2/3} \, \text{GeV}^2$, $Z_{nuc}=26$, $A_{nuc}=56$. 

Note that the formulas above are directly applicable to both scalar and pseudoscalar sgoldstino production. 

\subsection{Sgoldstinos from decaying mesons}
\label{Subsec:indirect}

{\it A priori} the most promising source of sgoldstino is long-lived mesons, for which the branchings into sgoldstino would be higher, than those of short-lived mesons. However, in case of ATLAS proton collision point the matter-free area for outgoing highly energetic mesons is rather restricted. Upon scattering off material the mesons lose energy and change the direction, significantly diminishing the chance to decay into sgoldstino which trajectory passes through the fiducial volume of the FASER detector. 

\begin{figure}[!htb]\center{
    \includegraphics[width=0.5\textwidth]{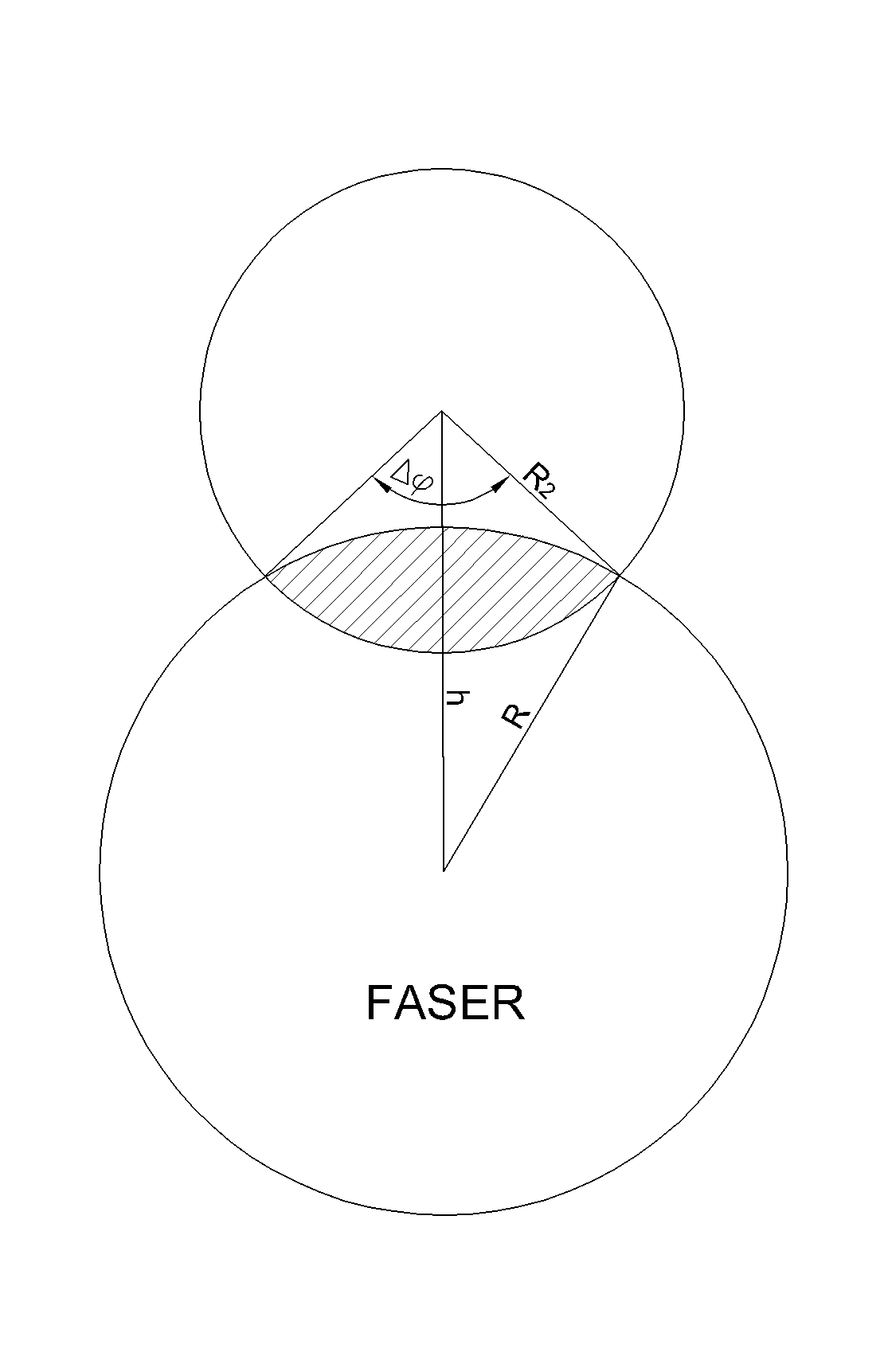}}
    \caption{Sgoldstino trajectory crossings (circle) the transverse to the beam line plane at the detector place (the encircled region  denoted as FASER). The signal region is shaded.}
    \label{raspad}
\end{figure}

We start with short-lived mesons. Let the parent meson be characterized in the laboratory frame (collision point at rest) by mass $m$ and transverse and longitudinal momenta $p_T$ and $p_L$, respectively. Then it is convenient to define some auxiliary kinematic quantities, such as meson transverse mass, rapidity, polar angle with respect to the collision axis and gamma-factors, in a standard way 
\begin{equation}
\begin{split}
    & m_T \equiv \sqrt{m^2+p_T^2}\;, \quad E \equiv m_T \cdot \cosh{y}\;, \quad p_L \equiv m_T \cdot \sinh{y}\;, \\
    & p \equiv \sqrt{E^2 - m^2}\;, \quad \gamma \equiv \frac{E}{m}\;, \quad \Gamma \equiv \frac{p}{m}\;, \quad \tan{\theta} \equiv \frac{p_T}{p_L}\;.
\end{split}
\end{equation}
And sgoldstino can be characterized by its 3-momentum $p_{S_0}$ and polar and azimuth angles $A$ and $\phi$ with respect to the polar axis along the parent meson 3-momentum. In particular, for the two-body decay into sgoldstino and massless particle, e.g. photon, one finds 
\begin{equation}
    p_{S_0} = \frac{m^2-m_S^2}{2m}\;, \quad E_{S_0} = \sqrt{p_{S_0}^2 + m_S^2}\;. 
\end{equation}
Then, making a boost back along the meson trajectory one obtains 
\begin{equation}
    p_{S1z} = \Gamma \cdot E_{S0} + \gamma \cdot p_{S0} \cos{A}\;, \quad E_{S1} = \Gamma \cdot p_{S0} \cos{A} + \gamma \cdot E_{S0}\;, 
\end{equation}    
\begin{equation}
    p_{S_{1_T}} =  p_{S_0} \cdot \sin{A}\;, \quad p_{S_1} = \sqrt{p_{S_{1_T}}^2 + p_{S_{1_z}}^2}\;,
\end{equation}  
\begin{equation}
    \tan \theta' \equiv \frac{p_{S_{1_T}}}{p_{S_z}}\;.
\end{equation}
These parameters refer to the coordinate system rotated with respect to the laboratory frame by angle $\theta$. Sgoldstino trajectory goes at angle $\theta'$ with respect to the polar axis of this system, the azimuth angle is arbitrary. 
Thus, at the place of detector, such sgoldstinos  cross the plane, which is transverse to the beam direction, through the circle shown in Fig.\,\ref{raspad}.
The segment lengths are proportional to the distance $L_{min}$ between the FASER detector and the collision point,  
\[
R_2=L_{min}\times \tan\theta'\;,\;\;\;\;h=L_{min}\times \tan\theta\;,
\]
and they are related to the azimuth angle $\Delta\varphi$ as follows 
\[
R_2\sin\frac{\Delta\varphi}{2} = \frac{2\sqrt{p(p-R)(p-R_2)(p-h)}}{h}\;,\;\;\;\text{where}\;\;\;\;\;2p\equiv R+R_2+h\,. 
\]
Sgoldstinos within this angle pass through the detector. If $\theta+\theta'<\Theta$, the azimuth angle remains unconstrained, since all the sgoldstinos pass through the detector. 
If $\theta-\theta'<\Theta$ all the sgoldstinos pass by the detector. 
The total number of sgoldstino events in FASER in the case of production by short-lived mesons (c.f.~\eqref{eq:direct_prod}) is
\begin{equation}\label{eq:short-lived}
N^{\rm sl}_S= L\times\int_{0}^{\sqrt{s}} dp \frac{d\sigma(pp\to X)}{dp} \int   f_X(p_T)dp_T \times P_p(p) \int\frac{d\text{Br}(X\to S)}{d p_S}dp_S\,.
\end{equation}

\begin{figure}[!htb]
\center{
    \includegraphics[width=0.7\linewidth]{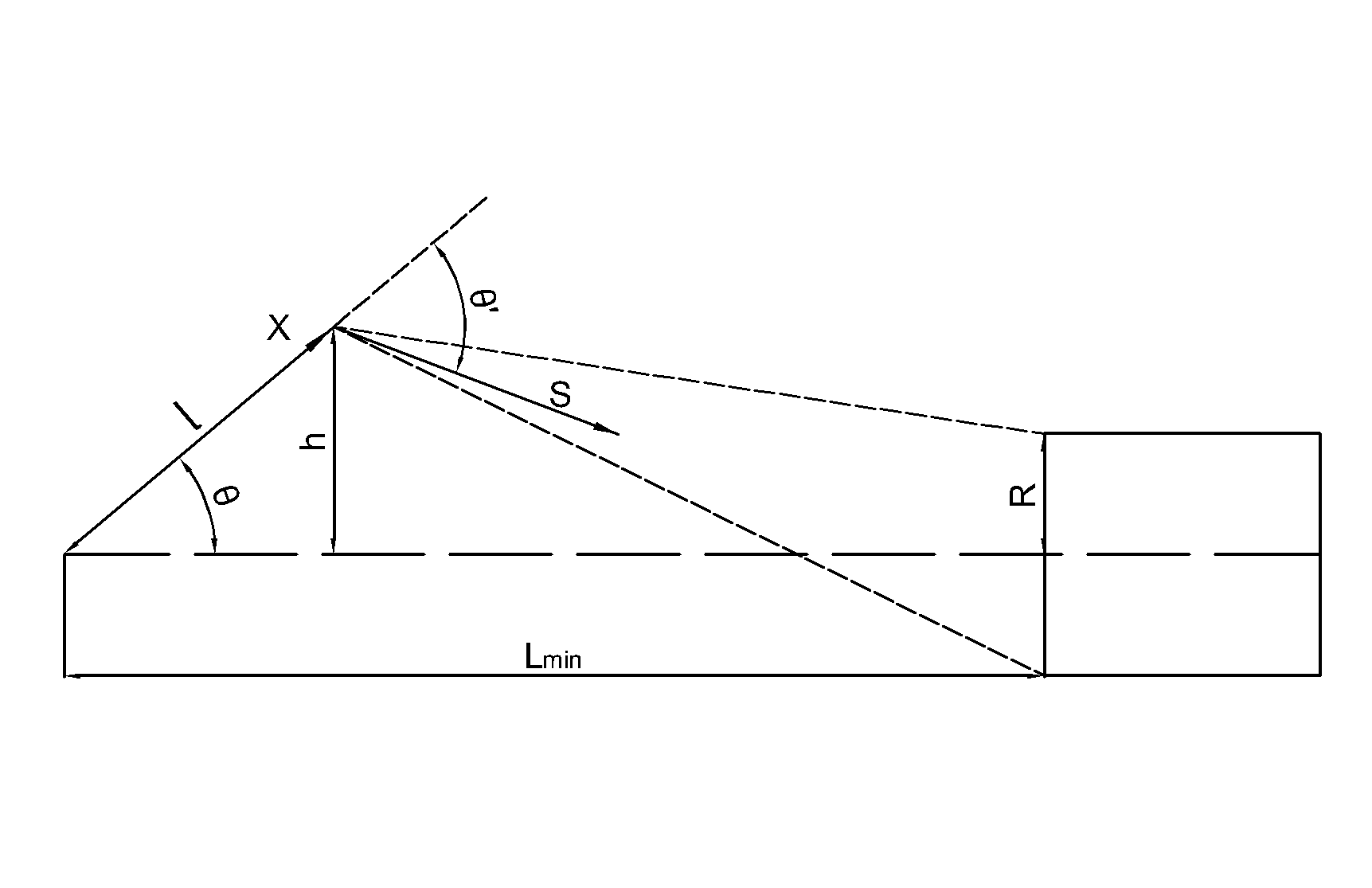} \\}
    \caption{Decay of long-lived meson $X$ into sgoldstino.}
    \label{fig:sketch}
\end{figure}

Then we consider the situation when sgoldstino is produced by a long-lived meson, which travels a distance $l$ before decay into sgoldstino. While the calculation of the sgoldstino 3-momentum in the laboratory frame goes along the same lines as in the previous case of the parent short-lived meson, the sgoldstino position at production is different. The suitable trajectories are illustrated with a sketch in Fig.\,\ref{fig:sketch}.
Only the sgoldstino trajectories between the short-dashed lines pass through the detector, the polar angle of sgoldstino 3-momentum are confined between 
\begin{equation}
    \theta'_{min}=\theta+\arctan \left( \frac{|h-R|}{L_{min}-l} \right),\;\;\;\;
    \theta'_{max}=\theta+\arctan \left( \frac{h+R}{L_{min}-l} \right)\,.
\end{equation}

\begin{figure}[!htb]
    \centering
    \includegraphics[width=0.9\textwidth]{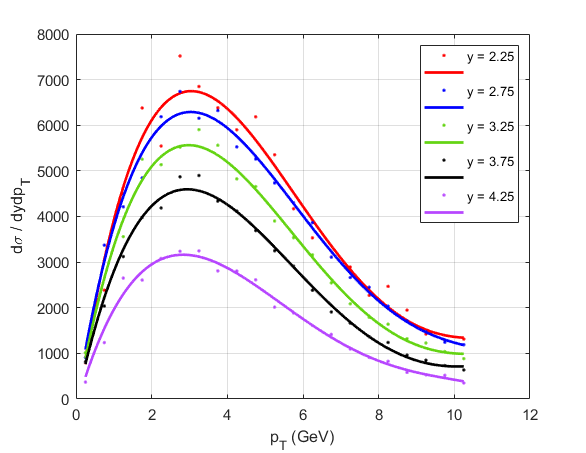}
    \caption{Approximation to the beauty meson differential cross section (over rapidity $y$ and transverse momentum $p_T$) we used to calculate the sgoldstino flux at FASER.}
    \label{dsdptB}
\end{figure}

The distance covered by the meson $X$ before decay (with lifetime $\tau_X$) is determined by the meson lifetime and 3-momentum $p$ (or $\gamma(p)$-factor). Namely, the probability for the meson to decay over the interval between $l$ and $l+d l$ reads
\[
dP_X(l,p)=\frac{ dl}{\tau_X \gamma}\text{e}^{-l/(\tau_X\gamma)}\,, 
\]
and for our case in Fig.\,\ref{fig:sketch} we have $l=h/\sin{\theta}$. 
The total number of sgoldstino events in FASER in the case of production by short-lived mesons (c.f.~\eqref{eq:direct_prod}) is
\begin{equation}\label{eq:long-lived}
N^{\rm ll}_S = L\times\int_{0}^{\sqrt{s}} dp \frac{d\sigma(pp\to X)}{dp} \int dP_X(l,p)\int   f_X(p_T)dp_T \times P_{np}(p,l)\int\frac{d\text{Br}(X\to S)}{d p_S}dp_S\,.
\end{equation}
The integrations over sgoldstino momenta in~\eqref{eq:short-lived} and~\eqref{eq:long-lived} are subject to the geometrical restriction (passing through the FASER detector volume) as we explained above.
Integration with the decay probability $P_X(l,p)$ goes along the  meson trajectory till it  comes to the absorbing matter (TAS or TAN, see FASER layout geometry for details\,\cite{fasercollaboration2019faser}.)  
The differential cross sections of meson $X$ production $d\sigma(pp\to X)/dp$ was taken from Refs.\,\cite{LHCb:2015foc,LHCb:2018yzj} for $X=J/\psi$ and $X=\Upsilon$. We approximate the beauty meson differential production cross section at LHC by fitting to the Tables of Ref.\,\cite{LHCb:2017vec}, the result is presented in Fig.\,\ref{dsdptB}. 
Similarly we obtain the distributions for other short-lived mesons. For kaons and $\eta$-mesons we utilized the simulations of $10^5$ proton-proton collisions with CMRC package and hadronic generator EPOS-LHC\,\cite{Pierog_2015,internetCRMC,HepMC2}. 

\section{FASER sensitivity to sgoldstino model parameters}
\label{Sec:sensitivity}

In previous sections we calculated sgoldstino flux in the FASER detector taking into account the direct production as well as production by short-lived and long-lived mesons.  In what follows, for concreteness, we fix two (least important for our study) parameters in the Higgs sector as  
\[
\mu=1\,\text{TeV}\,,\;\;\;\;\;\;\;m_A=1\,\text{TeV} 
\]
and scan over the major model parameters obeying the following natural constraints
\[
M_{\gamma\gamma},\,A_l,\,A_Q,\,M_3<\sqrt{F},\;\;\;\;\;\;3\,\text{TeV}< M_3,\;\;\;\;\;\;
m^{LR}_{ij}\le 100\,\text{GeV} \;\; \text{for} \;\; i \ne j
\]
to indicate the region which can be tested at the FASER experiment. 
To this end we assume zero background and set the number of signal events to be $N=3$ which implies a bound at 95\%\,CL according to the Poisson statistics. We also check, that constraints on rare meson decays from Tab.\,\ref{tab:tab4} are always satisfied. The two requirements --- to have as large production rate as possible and to keep the decay length at the level of 500\,m (as we explained above) --- shape the realistic region for the scanning procedure for each of the cases we investigate. It is worth noting, that the limit on the off-diagonal left-right squark mass terms is chosen to pass the constraints from the absence of rare processes even in models with relatively light squarks, about 3\,TeV. The heavier squarks give smaller contribution to the rare processes, and so the off-diagonal elements may be larger than 100\,GeV. We do not elaborate on this option here and discuss it in Conclusions. { In our calculations we choose some values for the set of soft supersymmetry breaking parameters and then scan over $\sqrt{F}$ and $m_S$. 
Wherever the scan crosses the unitarity border for a soft parameter $\sqrt{F}=M_{soft}$ we proceed to the region of lower $\sqrt{F}$ but with the soft term equal to $M_{soft}=\sqrt{F}$.}   

Since the production mechanisms at the FASER experiment are indistinguishable, we concentrate on the decay signatures instead, which for sgoldstino are decays into a pair of photons, charged leptons and mesons (three mesons for the pseudoscalar sgoldstinos) inside the FASER detector. Thus, we first consider all possible mechanisms which possibly lead to the  photon signature, then to the lepton signature and finally to the meson signature and for each mechanism and signature outline the model parameter space which can be explored at FASER. 
Namely, for each case we present several plots indicating the region in ($m_S(m_P)$, $\sqrt{F}$) plane for a particular set of other model parameters, and finally present the borders of the whole region to be potentially tested at FASER with a particular signature.  

{ Following previous studies devoted to new physics at FASER we assume zero background and 100\% efficiency for the detection of SM charged particles and photons. According to studies~\cite{Feng_2018,FASER:2018ceo,FASER:2018bac} the search for a pair of energetic ($\sim 1$~TeV scale) oppositely charged tracks can be performed with negligible background. As discussed in~\cite{FASER:2018ceo} the expected background of single photons with energy around TeV is very low, therefore even a pair of photons misreconstructed as a single shower represents a new physics signature. }

Here we still comment on general {characteristics} of different production mechanisms. In the case of direct sgoldstino production in proton-proton collisions the production rate grows with coupling to gluons
proportional to $M_3$, which simultaneously {raises} the decay rate into hadrons and shortens the sgoldstino lifetime. So naturally the only promising  case here is that of relatively light sgoldstinos. In the case of sgoldstino production in meson decays the production rate  is proportional to $A_Q$ rather than $M_3$, which can favor the models with heavier and sufficiently long-lived sgoldstinos. 

{ Before plunging into detailed discussion of FASER/FASER2 sensitivities to the model with light sgoldstino let us briefly comment on prospects of similar searches at FASER$\nu$ which is more recent proposal~\cite{FASER:2020gpr} whose main purpose is to detect high energy neutrinos produced at LHC. FASER$\nu$ detector of size $25$~cm$\times 25$~cm$\times 1.3$~m (and $40$~cm$\times 40$~cm$\times 8$~m for its successor FASER$\nu2$~\cite{Feng:2022inv}) is made of emulsion films interleaved with 1.1~tons (20 tons for FASER$\nu2$) of tungsten and it allows for  reconstructing tracks of charged particles with remarkable precision. FASER$\nu$ will perform data taking during run~III of LHC. It is straightforward to show that the probability of direct interactions of sgoldstino with mass around GeV scale in the FASER$\nu$ detector (before entering FASER) with production of charged particles is strongly suppressed as compared to the probability of its decay there (assuming the same interaction type). On the other hand, sgoldstino decays into a pair of charged particles inside FASER$\nu$ can be used to increase the sensitivity of FASER-I(II).}

\subsection{Signature: a pair of photons}
\label{Subsec:photons}
Since sgoldstino decays into photons are naturally suppressed if decays into mesons are kinematically allowed, the mode into photons are mostly promising for light sgoldstino, whose masses are below or not far from the threshold of the pion pair production. 

\begin{figure}[!htb]  
\centerline{
\includegraphics[width=0.5\textwidth]{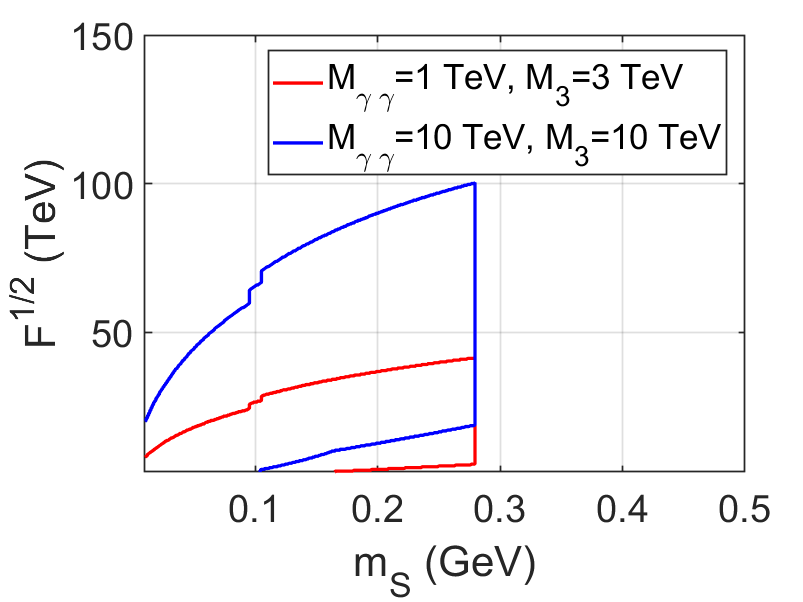}
\hskip 0.03\textwidth
\includegraphics[width=0.5\textwidth]{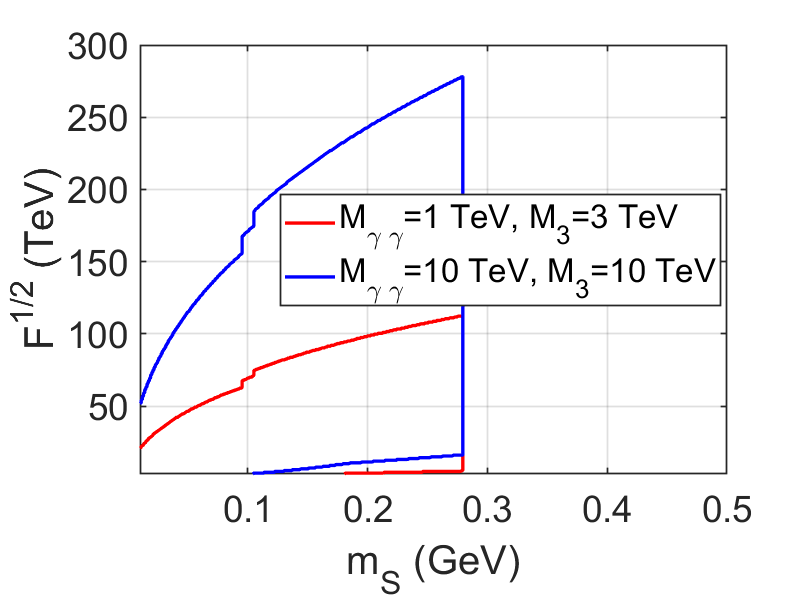}
}  
\caption{FASER-I (left plot) and FASER-II (right plot) sensitivity to models with light sgoldstino produced in gluon fusion and decaying into a couple of photons. { In the regions where $\sqrt{F}<M_{\gamma \gamma}$ we set $M_{\gamma \gamma}=\sqrt{F}$ and the same for $M_{3}$.}} 
\label{fig:PhotonFissionFASER}
\end{figure}

\paragraph{Direct production.} In this case  as we discussed above only relatively light sgoldstinos can be long-lived with their decay to photons being the dominant channel.  The regions of the model parameters to be tested at FASER are outlined in Fig.\,\ref{fig:PhotonFissionFASER}.
One observes from Fig.\,\ref{fig:PhotonFissionFASER} that higher $M_{\gamma\gamma}$ shifts the testable region to   
 larger values of $\sqrt{F}$ plainly to keep the decay length around the distance to the FASER detector, 500\,m. Higher $M_3$ simply increases the sgoldstino production cross section and hence inflates the region in all directions (still below the meson pair threshold). The step-like features in these and subsequent plots are related to jump discontinuities in $\beta$-function in \eqref{rg_factor}. Expectedly, the FASER-II is more sensitive, which is certainly due to its larger cross section and volume.     

\begin{figure}[!htb]  
\centerline{
\includegraphics[width=0.5\linewidth]{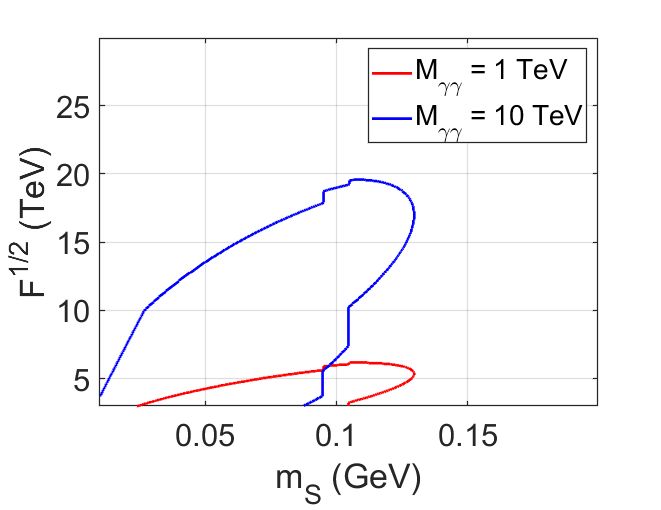}
\hskip 0.03\textwidth 
\includegraphics[width=0.5\linewidth]{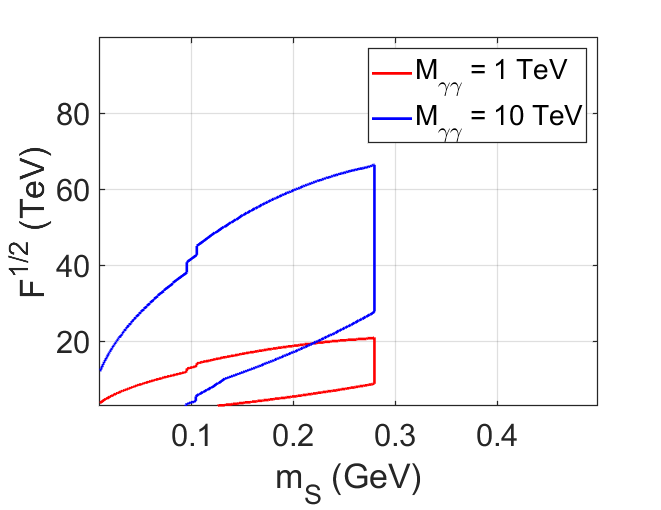} }  
\caption{Sensitivity region of FASER-I (left plot) and FASER-II (right plot) to models with scalar sgoldstino. Sgoldstinos are produced in Primakoff process and decay into a pair of photons. {In the regions where $\sqrt{F}<M_{\gamma \gamma}$ we set $M_{\gamma \gamma}=\sqrt{F}$}} 
\label{fig:PrimSFASER}
\end{figure}

\paragraph{Production in Primakoff process.}
Another promising source of sgoldstino is Primakoff process. Production rate is governed by $M_{\gamma\gamma}$ so the dominating mode of sgoldstino decay is into two photons.  The corresponding sensitivity regions of FASER are presented in Fig.\,\ref{fig:PrimSFASER}. 
Here, similar to the case of direct production, the signal regions in sgoldstino masses are  limited by the decay threshold into pions. For a realistic sets of model parameters this production mechanism is promising only for the two photon signature. 

\begin{figure}[!htb]  
\centerline{
\includegraphics[width=0.5\textwidth]{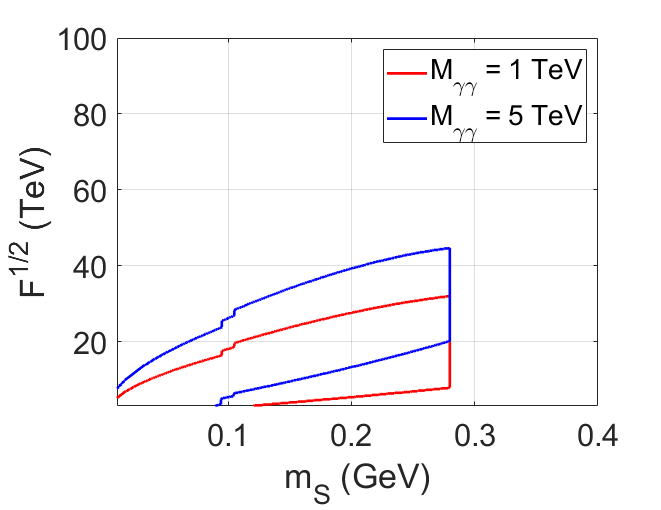} 
\hskip 0.03\textwidth
\includegraphics[width=0.5\textwidth]{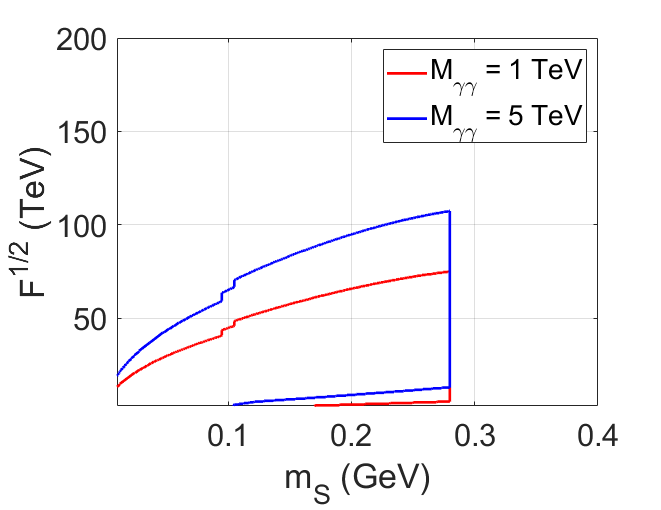}
}  
\caption{Sensitivity regions of FASER-I (left panel) and FASER-II (right panel). Sgoldstinos  are produced in $\eta$-meson decays and decay inside the detector into a couple of photons. $A_Q = 100$ GeV, $M_3=3$ TeV. {In the regions where $\sqrt{F}<M_{\gamma \gamma}$ we set $M_{\gamma \gamma}=\sqrt{F}$.}} \label{fig:PhotonCons1FASER}
\end{figure}

\begin{figure}[!htb]  
\centerline{
\includegraphics[width=0.5\textwidth]{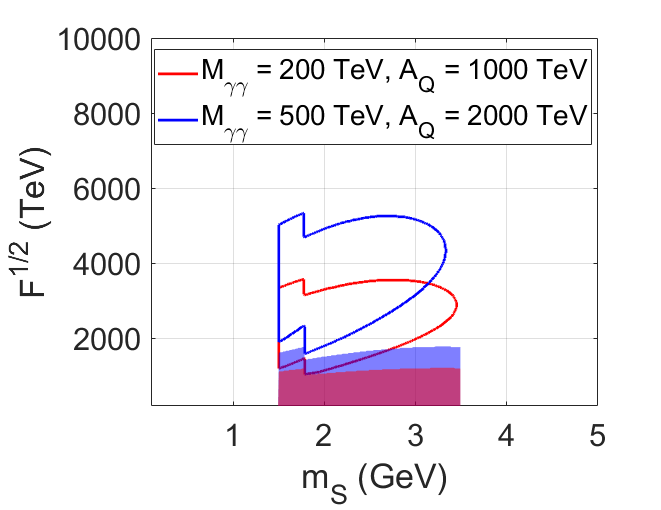}
}
 \caption{FASER-II sensitivity to the model parameter space, where sgoldstino is produced in beauty meson decays and decay into a photon pair within the detector. $M_3 = 3$ TeV. { Shaded areas correspond to restrictions from tab.\,\ref{tab:tab4}. In the regions where $\sqrt{F}<A_Q$ we set $A_Q=\sqrt{F}$.}} 
\label{fig:PhotonCons2FASER}
\end{figure}

\paragraph{Production in meson decays. Flavor conserving sgoldstino couplings.} In this case there is a chance to make the photon mode promising for heavier sgoldstinos in the models with $M_{\gamma\gamma}\gg M_3$. Light sgoldstinos are mostly produced in decays of $\eta$-mesons, the corresponding sensitivity regions are shown in Fig.\,\ref{fig:PhotonCons1FASER}.
Similar contribution of $\eta'$-mesons are negligible, since their production rate in proton-proton scattering is much lower than that of $\eta$-mesons. At higher $\sqrt{F}$ decays of $B$-mesons contribute as well. Remarkably, heavier sgoldstinos, up to kaon pair threshold, may become visible at FASER-II, see Fig.\,\ref{fig:PhotonCons2FASER}. 
We found no room to probe this case at the first stage of FASER operation.  

\begin{figure}[!htb]  
\centerline{
\includegraphics[width=0.5\textwidth]{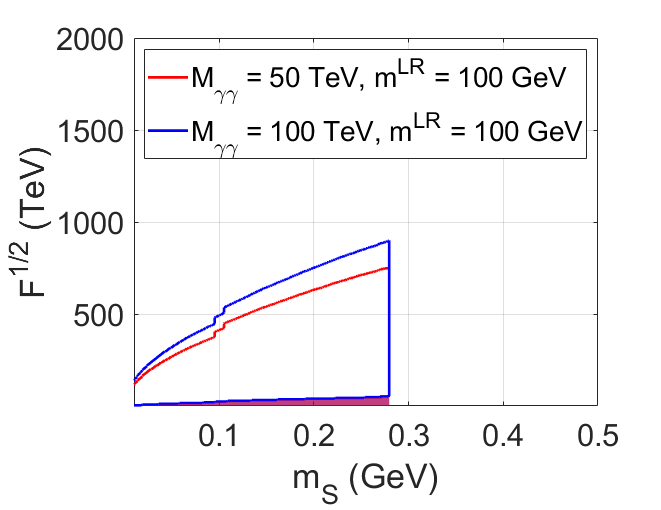}}  
\caption{Sensitivity regions of FASER-I (photon signature) to models with sgoldstino production via $\eta'$ decays due to sgoldstino flavor-violating coupling to quarks. {Shaded areas correspond to restrictions from tab.\,\ref{tab:tab4}. In the regions where $\sqrt{F}<M_{\gamma \gamma}$ we set $M_{\gamma \gamma}=\sqrt{F}$.} } 
\label{fig:PhotonVFASER1}
\end{figure}

\begin{figure}[!htb]  
\centerline{
\includegraphics[width=0.5\linewidth]{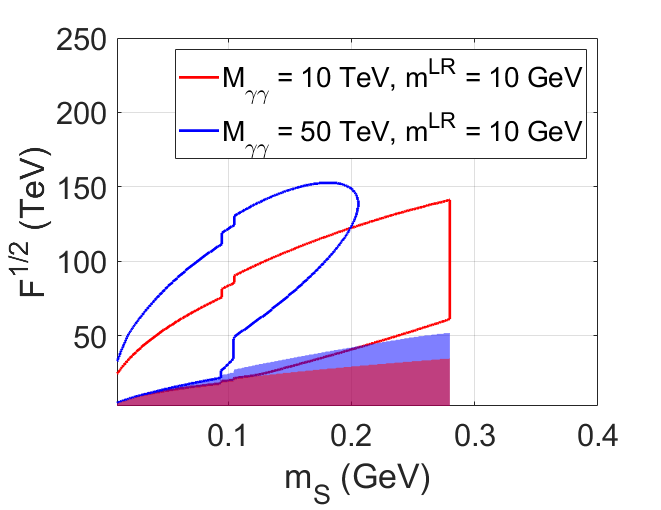}
\hskip 0.03\textwidth 
\includegraphics[width=0.5\linewidth]{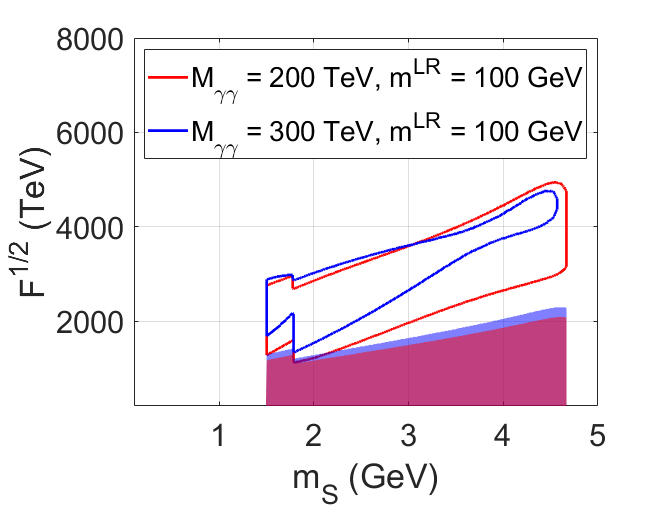} }  
\caption{Sensitivity region of FASER-II to models with sgoldstino flavor-violating couplings to quarks. Sgoldstinos are produced in decays of charm mesons (left panel) and beauty mesons (right panel). {Shaded areas correspond to restrictions from tab.\,\ref{tab:tab4}. At points where $\sqrt{F}<M_{\gamma \gamma}$ we take $M_{\gamma \gamma}=\sqrt{F}$.} } 
\label{fig:PhotonV1FASER}
\end{figure}

\paragraph{Production in meson decays. Flavor violating sgoldstino couplings.} Flavor-violating sgoldstino couplings provide with another alternative mechanism to have a reasonably high sgoldstino production rate while {keeping sgoldstino}  sufficiently long-lived and dominantly decaying to photons. For the first stage of FASER operation only light sgoldstinos can be tested in this way. The dominant production is due to $\eta'$ {and K}-meson decay, see the relevant regions in Fig.\,\ref{fig:PhotonVFASER1}. 
For the second stage, FASER-II, the heavy mesons can contribute as well.  
Remarkably, sgoldstinos much heavier than the pion threshold can be probed being produced  mostly in beauty meson decays, see Fig.\,\ref{fig:PhotonV1FASER}. 

To conclude on the photon pair signature, we observe that it is rather generic in models with sgoldstinos lighter than the pion pair threshold, but for a specific pattern of MSSM soft terms can be promising in testing even GeV-scale sgoldstinos. Investigating separately specific mass intervals we found that FASER-I can explore the following broad regions of the model parameter space 
\begin{equation}
\begin{split}
    & \sqrt{F} < 32 \cdot 10^3 \, \text{TeV} \\
    & M_{\gamma \gamma}< \sqrt{F} \\
    & M_3< \sqrt{F} \\
    & A_Q< \sqrt{F} \\
    & m_{ij}^{LR}<100 \, \text{GeV}
\end{split}
\end{equation}
for $m_S<2m_\pi$. 

The FASER-II prospects are expectedly brighter and the regions in the sgoldstino model parameter space are correspondingly wider:  
\begin{equation}
\begin{split}
    & \sqrt{F} < 125 \cdot 10^3 \, \text{TeV} \\
    & M_{\gamma \gamma}< \sqrt{F} \\
    & M_3<\sqrt{F} \\
    & A_Q<\sqrt{F} \\
    & m_{ij}^{LR}<100 \, \text{GeV}
\end{split}
\end{equation}
for $m_S<2m_\pi$ and 
\begin{equation}
\begin{split}
    & \sqrt{F} < 25 \cdot 10^3 \, \text{TeV} \\
    & M_{\gamma \gamma}< \sqrt{F} \\
    & A_Q< \sqrt{F} \\
    & m_{ij}^{LR} < 100 \, \text{GeV}
\end{split}
\end{equation}
for $1.5 \, \text{GeV} < m_S < 4.8$\,GeV. 

\subsection{Signature: a couple of charged leptons}
\label{Subsec:leptons}

Sgoldstino decay rates into fermions are proportional to the fermion mass squared $m_f^2$, see eq.\,\eqref{eq22}. Hence, it is  naturally suppressed with respect to decay into a pair of vectors (photons or gluons) far from the fermion pair threshold region $m_S\approx 2m_f$. The electron-positron decay mode is naturally interesting only for very light sgoldstinos, $m_S\lesssim 10$\,MeV, which is beyond the scope of this paper. Decay mode into muon pair can be interesting for $m_S\sim 0.2-4.8$\,GeV, and especially for the narrow mass range  $2m_\mu<m_S<2m_\pi$, where it can naturally dominate. In what follows we consider only decays into muon pairs. 

\begin{figure}[!htb]  
\centerline{
\includegraphics[width=0.5\textwidth]{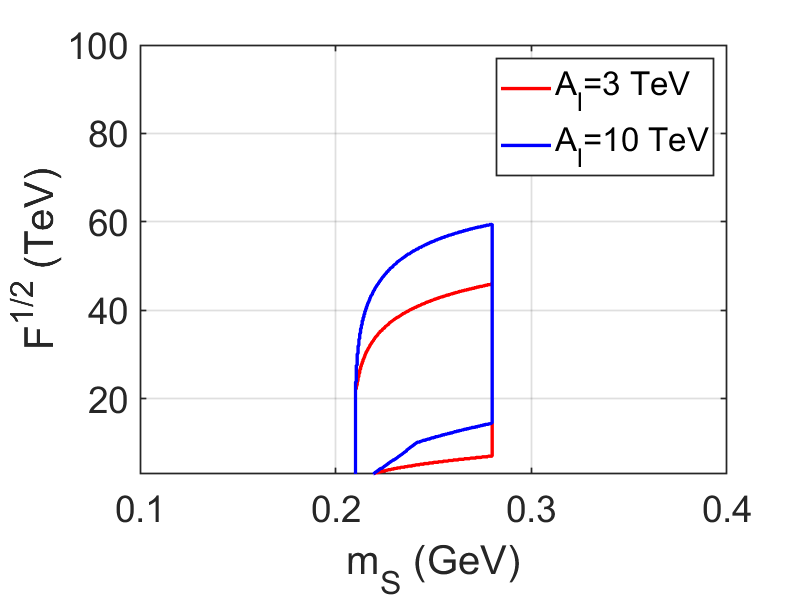}
\hskip 0.03\textwidth \includegraphics[width=0.5\textwidth]{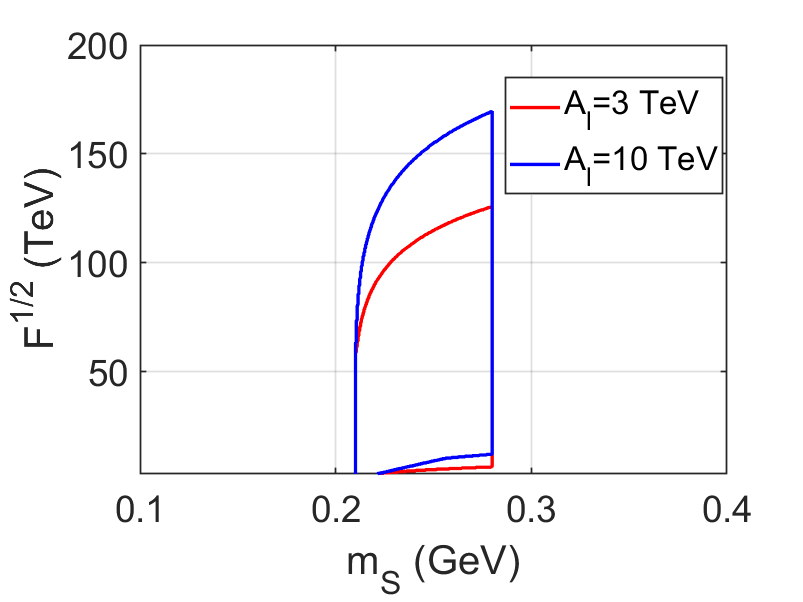}}
\caption{The regions to be tested with the muon pair signature at FASER-I (left panel) and FASER-II (right panel). $M_3$ = 3 TeV. {In the regions where $\sqrt{F}<A_l$ we set $A_l=\sqrt{F}$.} } 
\label{fig:LeptonFASER}
\end{figure}

\begin{figure}[!htb]  
\centerline{
\includegraphics[width=0.5\textwidth]{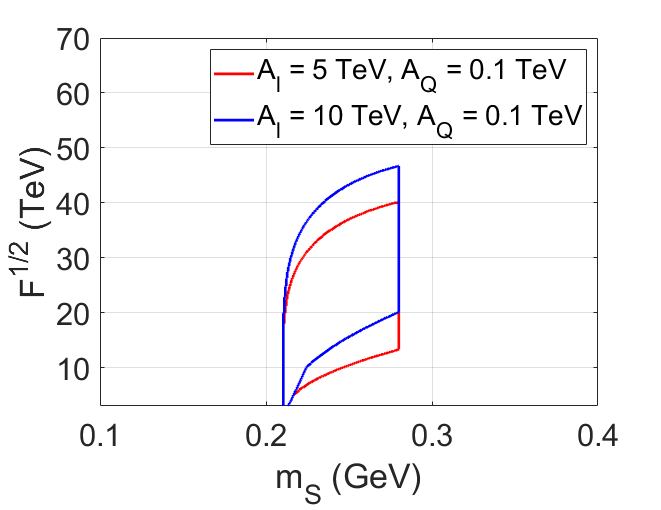} 
\hskip 0.03\textwidth 
\includegraphics[width=0.5\textwidth]{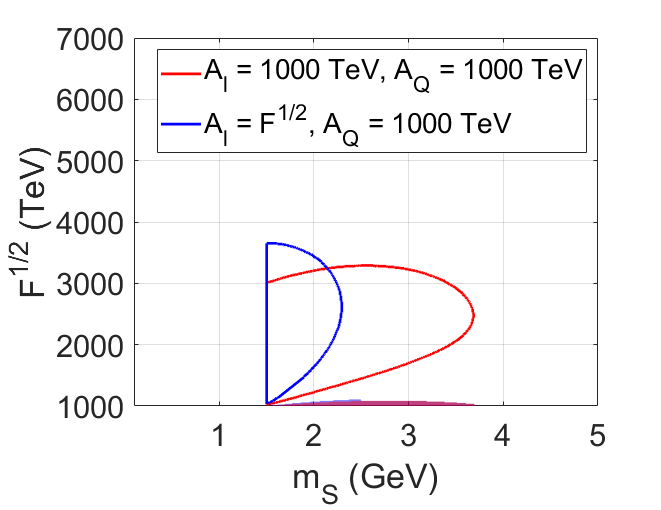} } 
\caption{The sensitivity regions of FASER-I to sgoldstino, produced in $\eta-$meson decays with $A_Q = 100$ GeV (left panel) and FASER-II to sgoldstino, produced in beauty meson decays (right panel) due to flavor-conserving couplings, and decaying to a  couple of muons inside the detector. $M_3=3$ TeV, $M_{\gamma\gamma} = 100$ GeV. {Shaded areas correspond to restrictions from tab.\,\ref{tab:tab4}. In the regions where $\sqrt{F}<A_l$ we set $A_l=\sqrt{F}$.} } 
\label{fig:LeptonConsFASER}
\end{figure}

\paragraph{Direct production.} Sgoldstinos can be tested only in a very narrow mass interval, see Fig.\,\ref{fig:LeptonFASER}.
The room for FASER-II is wider, as for all the other modes. The changes of the region with trilinear coupling can be explained similar to the case of photon pair signature and the region changes with $M_{\gamma\gamma}$ as we discuss in Sec.\,\ref{Subsec:photons}. Namely: increasing the value of trilinear coupling one increases the sgoldstino-decay rate, and so must increase the scale of supersymmetry breaking $\sqrt{F}$ to keep the sgoldstino life-time in the optimal region to decay inside the FASER detector.  
\paragraph{Production in meson decays. Flavor conserving sgoldstino couplings.} Here for 
the first stage of FASER only $\eta$-meson decays can give a noticeable contribution to the sgoldstino production. For heavier sgoldstinos the pion decay modes make muon channel irrelevant. 
At FASER-II heavy sgoldstinos could be produced in beauty meson decays with large enough $\sqrt{F}$ and $A_l$ so the lepton decay mode dominates. The corresponding region to be explored there is presented in Fig.\,\ref{fig:LeptonConsFASER}.

\begin{figure}[!htb]  
\centerline{
\includegraphics[width=0.5\textwidth]{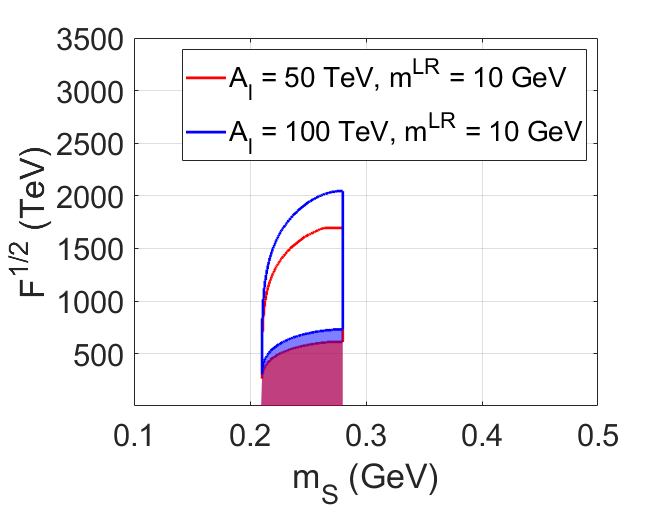}
\hskip 0.03\textwidth 
\includegraphics[width=0.5\textwidth]{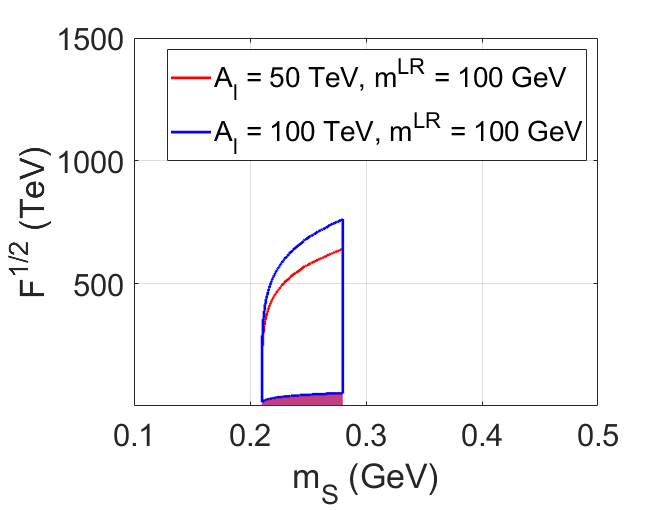}
}
\caption{FASER-I sensitivity to light sgoldstinos produced in K-mesons, where we take $m_{D12}^{LR} = m_{D21}^{LR} = m^{LR} = 10$ GeV (left plot) and $\eta'$-meson (right plot) decays due to flavor violating couplings. Sgoldstino decays into muon pair inside the detector volume. 
{Shaded areas correspond to restrictions from tab.\,\ref{tab:tab4}. In the regions where $\sqrt{F}<A_l$ we set $A_l=\sqrt{F}$.} } 
\label{fig:LeptonV3FASER}
\end{figure}

\paragraph{Production in meson decays. Flavor violating sgoldstino couplings.} Only $\eta'-$ and K-- mesons contribute to the sgoldstino production in this case at FASER-I. The region available for study is illustrated with Fig.\,\ref{fig:LeptonV3FASER}. 

\begin{figure}[!htb]  
\centerline{ 
\includegraphics[width=0.5\textwidth]{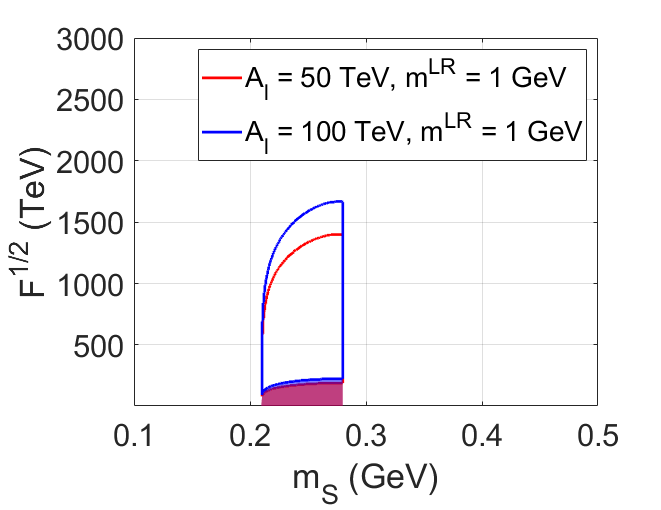} 
\hskip 0.03\textwidth 
\includegraphics[width=0.5\textwidth]{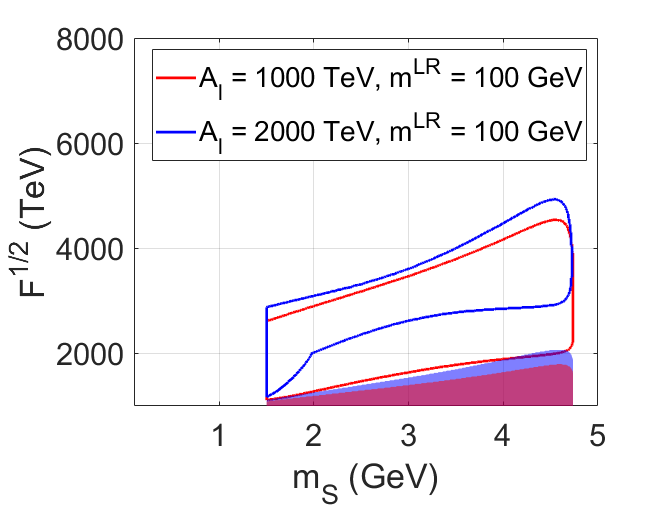} 
}
\caption{The region of FASER-II sensitivity to sgoldstinos produced in kaon decays, where we take $m_{D12}^{LR} = m_{D21}^{LR} = m^{LR} = 1$ GeV (left panel) and beauty meson decays (right panel) due to sgoldstino flavor-violating couplings. Sgoldstino decays into muon pair inside the detector. { Shaded areas correspond to restrictions from tab.\,\ref{tab:tab4}. In the regions where $\sqrt{F}<A_l$ we set $A_l=\sqrt{F}$.} } 
\label{fig:LeptonV1FASER2}
\end{figure}

The FASER-II prospects are richer: heavier sgoldstinos could be produced in beauty meson decays. The relevant region of the model parameter space is shown in Fig.\,\ref{fig:LeptonV1FASER2}. 

To summarize the two lepton pair signature, we found that FASER-I can test the following region 
\begin{equation}
\begin{split}
    & \sqrt{F} < 25 \cdot 10^3 \, \text{TeV} \\
    & 3 \, \text{TeV} < A_l< \sqrt{F} \\
    & M_3< \sqrt{F} \\
    & m_{ij}^{LR} < 100 \, \text{GeV}
\end{split}
\end{equation}
and only for $2m_\mu<m_S<2m_\pi$, while FASER-II can probe 
\begin{equation}
\begin{split}
    & \sqrt{F} < 92 \cdot 10^3 \, \text{TeV} \\
    & 3 \, \text{TeV} < A_l< \sqrt{F} \\
    & M_3< \sqrt{F} \\
    & A_Q< \sqrt{F} \\
    & m_{ij}^{LR} < 100 \, \text{GeV}
\end{split}
\end{equation}
for $m_S<2m_\pi$ and 
\begin{equation}
\begin{split}
    & 2 \cdot 10^3 \, \text{TeV}<\sqrt{F} < 14 \cdot 10^3 \, \text{TeV} \\
    & 2\cdot 10^3 \, \text{TeV} < A_l< \sqrt{F} \\
    & A_Q< \sqrt{F} \\
    & m_{ij}^{LR} < 100 \, \text{GeV}
\end{split}
\end{equation}
for heavier sgoldstinos, $1.5 \, \text{GeV} < m_S < 4.8$\,GeV. 

\subsection{Signature: two mesons}
\label{Subsec:mesons} 

In this case sgoldstino must be heavier than the meson pair threshold. In the model, where all the soft terms are of the same order, sgoldstinos then decay mostly into mesons, with photon  and lepton pairs being only subdominant. 

\begin{figure}[!htb]  
\centerline{
\includegraphics[width=0.5\textwidth]{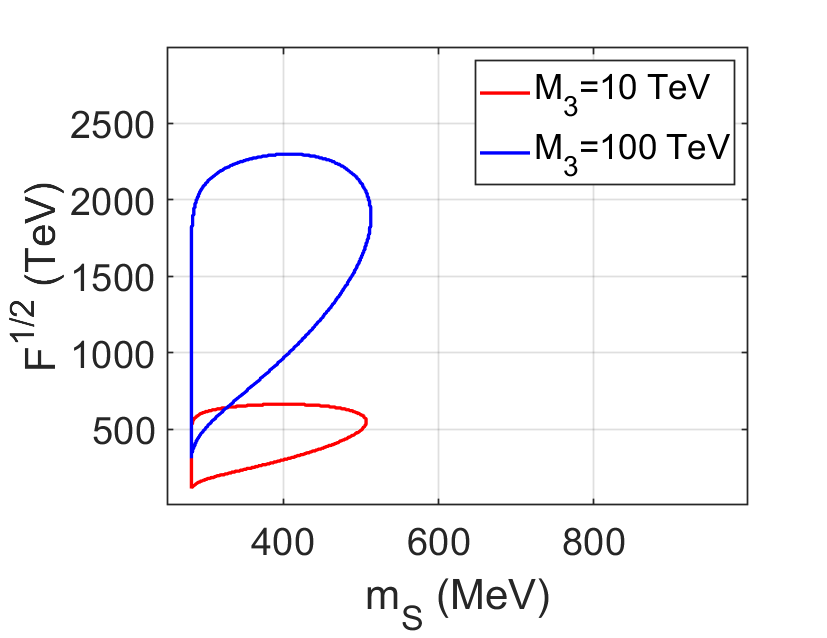} 
\hskip 0.03\textwidth 
\includegraphics[width=0.5\textwidth]{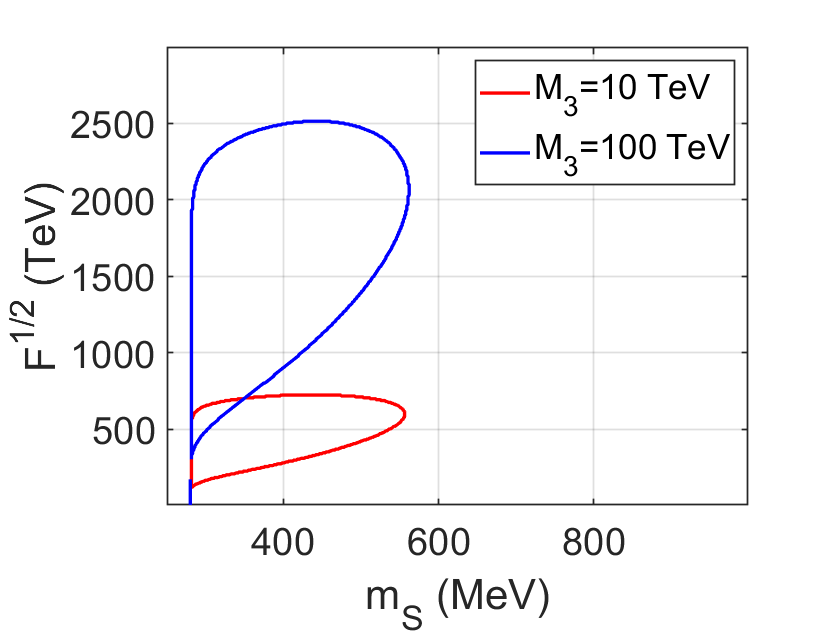} 
}
\caption{Sensitivity regions of FASER-II to the model parameters, where sgoldstinos are produced directly and then decay inside the detector volume into mesons. When only charged mesons are used as a signature, the regions are presented on left panel. The right panel shows the region, when both charged and neutral mesons are exploited as sgoldstino signatures. Only pions are kinematically viable. {In the regions where $\sqrt{F}<M_3$ we set $M_3=\sqrt{F}$.}} 
\label{MesonFissionFASER2}
\end{figure}

\paragraph{Direct production.} In this case both sgoldstino production and dominant decay rates are proportional to one and the same soft parameter, gluino mass squared $M_3^2$. An increase in the lifetime is compensated by decrease in the sgoldstino production. It happens then, that FASER-I has {no chance} to  explore this case. The viable examples of the region to be tested at the FASER-II are presented in Fig.\,\ref{MesonFissionFASER2}, 
considering separately signatures of only charged and of all light mesons. Kinematically, only pions are suitable. One observes, that adding the neutral meson decays to the sgoldstino signature somewhat enlarge the region of the model parameter space available for investigation with FASER-II detector. 

\begin{figure}[!htb]  
\centerline{
\includegraphics[width=0.5\linewidth]{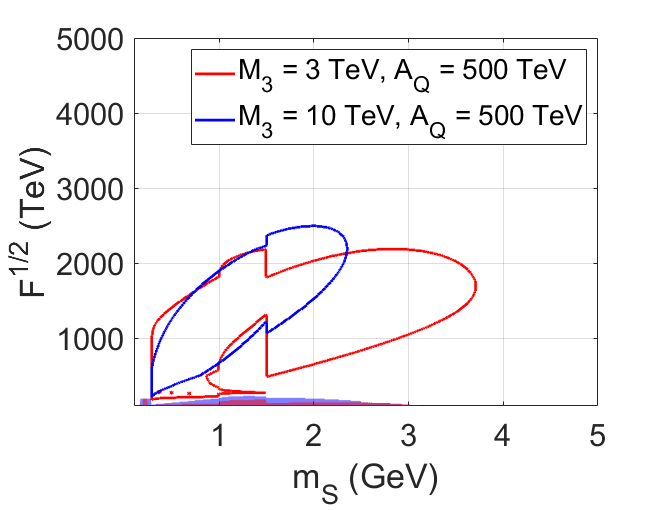} 
\hskip 0.03\textwidth 
\includegraphics[width=0.5\linewidth]{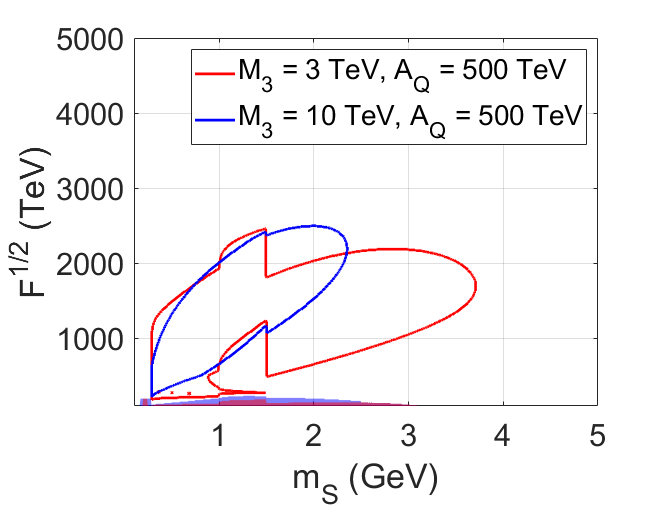} 
}
\caption{The regions to be tested with FASER-II detector. Sgoldstinos are produced by  $B$-mesons and decay into light mesons (pions and kaons) if $m_S<1.5\,$GeV and into pair of gluons if $m_S>1.5\,$ GeV. Only charged mesons are accounted for the left plot. Both charged and neutral meson are treated as viable signatures for the right plot. $M_{\gamma\gamma}=100$ GeV. {Shaded areas correspond to restrictions from tab.\,\ref{tab:tab4}. In the regions where $\sqrt{F}<A_Q$ we set $A_Q=\sqrt{F}$.}} 
\label{fig:MesonConsFASER2}
\end{figure}

\paragraph{Production in meson decays. Flavor conserving sgoldstino couplings.} The relevant examples of the regions in the model parameter space available for investigations there are outlined in Fig.\,\ref{fig:MesonConsFASER2}.  
Sgoldstinos are mostly produced by decaying beauty mesons, other contributions are negligibly small.

\begin{figure}[!htb]  
\centerline{
\includegraphics[width=0.5\textwidth]{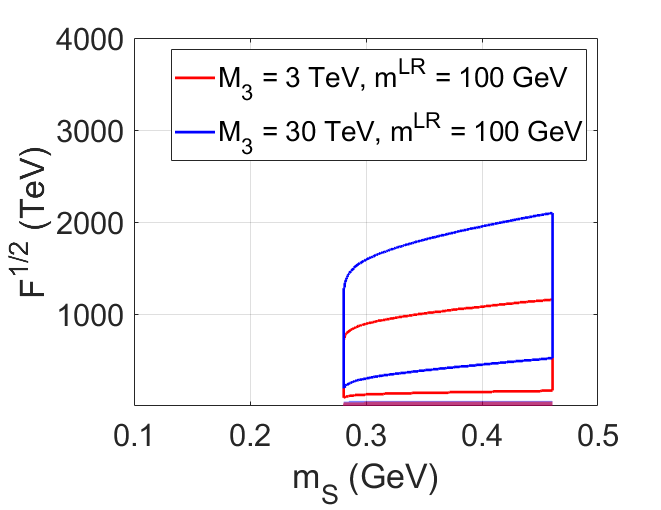} 
\hskip 0.03\textwidth  
\includegraphics[width=0.5\textwidth]{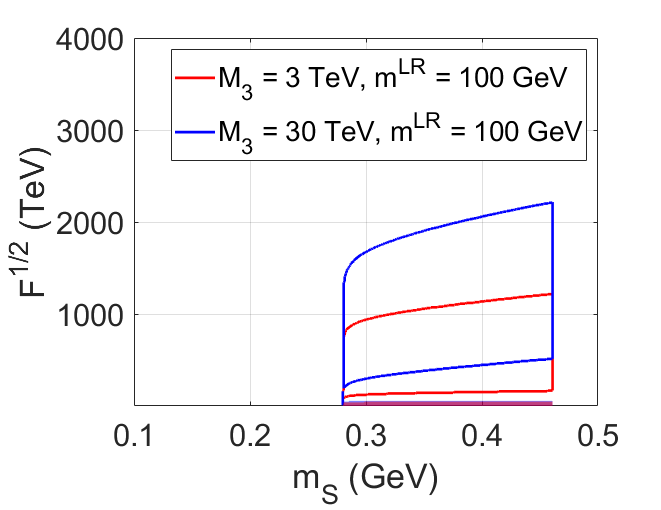} }
\caption{FASER-I sensitivity to models with light sgoldstino decaying into mesons. Sgoldstinos are produced mostly in flavor-violating decays of $\eta'$-mesons. Only sgoldstino decays into charged mesons are considered for the plot. Both neutral and charged mesons are treated as sgoldstino signature for the right plot. $A_Q=100$ GeV. { Shaded areas correspond to restrictions from tab.\,\ref{tab:tab4}. In the regions where $\sqrt{F}<M_3$ we set $M_3=\sqrt{F}$.} } \label{fig:MesonetaVFASER1}
\end{figure}

\paragraph{Production in meson decays. Flavor violating sgoldstino couplings.} In this case the first stage of FASER has a chance to probe the sgoldstino model. Then sgoldstinos are produced by flavor-violating decays of $\eta'$- and B-mesons. The available for testing regions are illustrated with Fig.\,\ref{fig:MesonetaVFASER1}.  

\begin{figure}[!htb]  
\centerline{
\includegraphics[width=0.5\textwidth]{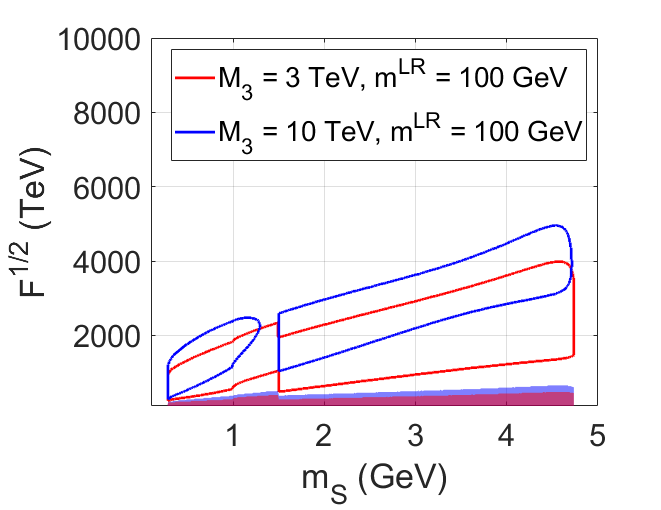} 
\hskip 0.03\textwidth \includegraphics[width=0.5\textwidth]{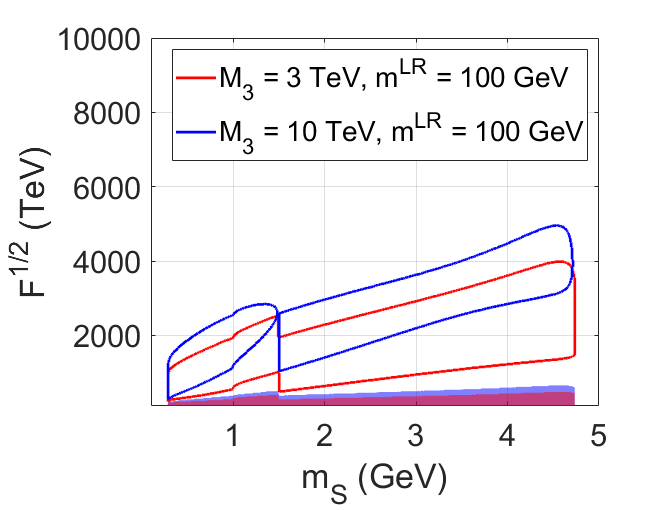} 
}
\caption{Regions of sgoldstino model parameters to be explored in FASER-II experiment. Sgoldstinos are mostly produced in beauty meson decays and exhibit signatures of a pair of charged mesons (left plot) and both neutral and charged mesons (right plot) for $m_S < 1.5\,$GeV. Sgoldstinos with $m_S>1.5\,$GeV decay into gluon pairs. $A_Q=100$ GeV. { Shaded areas correspond to restrictions from tab.\,\ref{tab:tab4}.}} \label{fig:MesonVFASER2}
\end{figure}

At the second stage of its operation, FASER can be sensitive to wider region of model parameter space. The examples of the viable regions in the model parameter space are shown in Figs.\,\ref{fig:MesonVFASER2}. 

Note, that the signal regions on these plots stretch to sgoldstino masses up to about 5\,GeV, where our assumption of two-meson dominance in the hadronic decay modes of sgoldstino may be invalid, and the main hadronic signature may be  decays into multiple pion states instead. However, this signature seems to be even more noticeable than the meson pair, so the hadronic modes still are promising in testing the supersymmetric models with light sgoldstinos.

In conclusion we see that with the pion pair signature of sgoldstino decay the FASER-I can test the following region of the model parameter space 
\begin{equation}
\begin{split}
    & \sqrt{F} < 3500 \, \text{TeV} \\
    & M_3< \sqrt{F} \\
    & 10 \, \text{GeV} < m_{ij}^{LR}<100 \, \text{GeV}
\end{split}
\end{equation}
The study is relevant for  $2m_\pi < m_S<0.46$\,GeV. On the other hand, FASER-II can explore much wider region, 
\begin{equation}
\begin{split}
    & \sqrt{F} < 25 \cdot 10^3 \, \text{TeV} \\
    & M_3<\sqrt{F} \\
    & A_Q<\sqrt{F} \\
    & m_{ij}^{LR}<100 \, \text{GeV}
\end{split}
\end{equation}
Likewise, this investigation is relevant for $2m_\pi < m_S < 4.8$\,GeV.

\subsection{Sensitivity to pseudoscalar sgoldstinos: two photons, two leptons and three mesons as promising signatures.}
\label{Subsec:pseudoscalar}
Here we repeat our study for pseudoscalar sgoldstino $P$. In several points this case is very similar to that of the scalar sgoldstino, e.g. decays into photons and leptons, direct production. However, decays into mesons and production in meson decays differ from the scalar case. Consequently, the regions of the model parameters which FASER can probe, if pseudoscalar sgoldstino is light, are not the same as the regions in models with scalar sgoldstino of the same mass. Below we present our results for each of the interesting signatures.  

\begin{figure}[!htb]  
\centerline{
\includegraphics[width=0.5\textwidth]{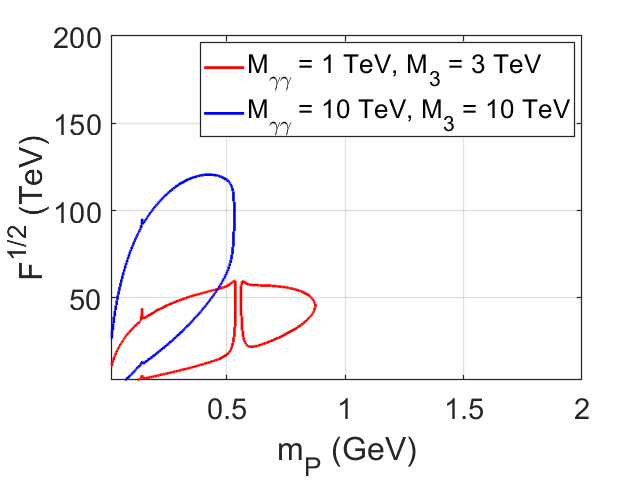}
\hskip 0.03\textwidth
\includegraphics[width=0.5\textwidth]{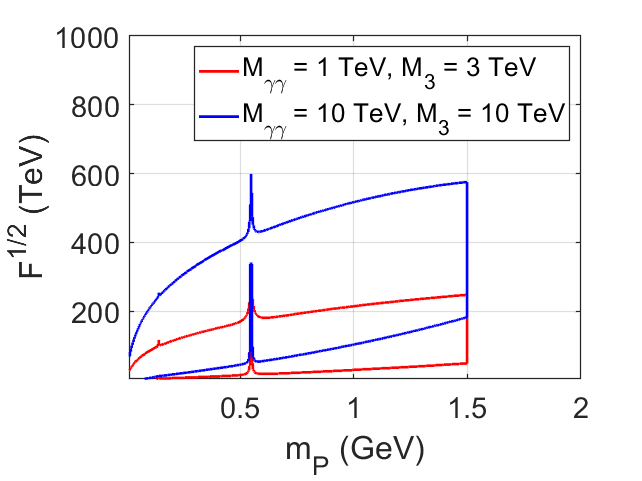}
}  
\caption{FASER-I (left plot) and FASER-II (right plot) sensitivities to models with light pseudoscalar sgoldstino produced in gluon fusion and decaying into a couple of photons. { In the regions where $\sqrt{F}<M_{\gamma \gamma}$ we set $M_{\gamma \gamma}=\sqrt{F}$ and the same for $M_3$.}} 
\label{fig:PhotonFissionPFASER}
\end{figure}

\paragraph{Two photons.} This signature for direct production mechanism is more promising here than in case of scalar sgoldstino for relatively low mass range. Indeed, for the same value of the gluon coupling constant, responsible for the production 
via gluon fusion, the lifetime of light pseudoscalar sgoldstino is longer than that of the scalar sgoldstino, since the allowed hadronic modes start with three pions and hence suppressed  as compared to two-body hadronic decays of the scalar. Long-lived particle has better chance to reach the detector placed at large distance from the production point, which is the case of FASER. The regions to be probed at FASER-I and FASER-II are outlined on left and right panels of Fig.\,\ref{fig:PhotonFissionPFASER} 
respectively. They are much wider than those in case of scalar sgoldstino, see Fig.\,\ref{fig:PhotonFissionFASER}. Pecularities at $m_S \approx 130$ MeV and  $m_S \approx 550$ MeV correspond to resonances in eq. (\ref{eq-pseudoscalarphotonpi}) and  eq.(\ref{eq-pseudoscalarphotoneta}).

\begin{figure}[!htb]  
\centerline{
\includegraphics[width=0.5\linewidth]{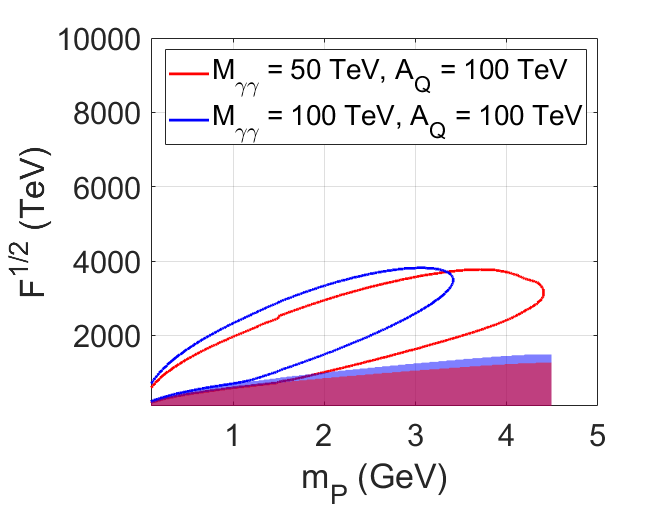}
\hskip 0.03\textwidth 
\includegraphics[width=0.5\linewidth]{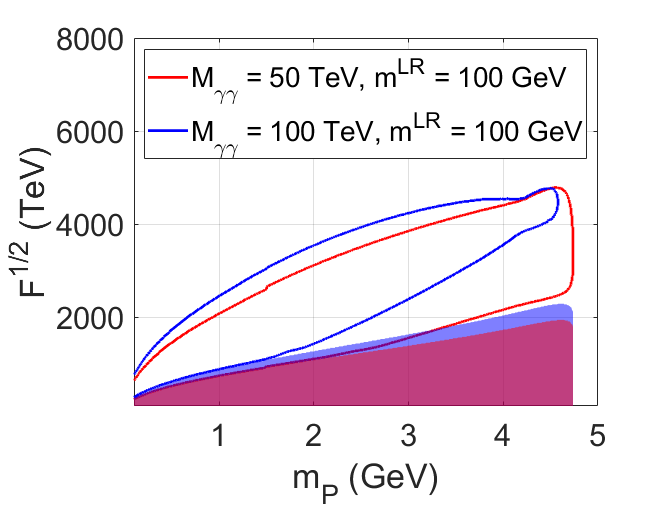} }  
\caption{Sensitivity region of FASER-II to models with sgoldstino  flavor-conserving (left panel) and flavor-violating (right panel) couplings to quarks. Pseudoscalar sgoldstinos are produced in decays of beauty mesons and decay into a pair of photons. $M_3=3$ TeV. { Shaded areas correspond to restrictions from tab.\,\ref{tab:tab4}.} } 
\label{fig:PhotonMesonPFASER}
\end{figure}

Quite the opposite, the pseudoscalar sgoldstino production in meson decays is generally less promising than that of scalar sgoldstino. Indeed, production in $\eta$-meson decays (which proceeds with three particles in the final state) is kinematically open only for $m_P<300$\,MeV and strongly suppressed by the phase space volume. We neglect this channel for the interesting range of  sgoldstino mass. Pseudoscalar sgoldstino may be produced in decays of heavy pseudoscalar meson with vector meson in the final state. Since vector mesons are excited states of the pseudoscalar mesons, they are naturally heavier, and so the kinematics shrinks somewhat the allowed mass range for the pseudoscalar sgoldstino with respect to what we have for the scalar sgoldstino. We considered both cases: when pseudoscalar sgoldstinos are produced in beauty meson decays to vector mesons and pseudoscalar mesons. Finally, while for $M_P\lesssim 1.5$\,GeV pseudoscalar sgoldstino  decays into hadrons (mesons) are naturally suppressed, while for heavier particles it is similar to that of scalar sgoldstino and is saturated by decays into gluons, which then hadronize. Thus we found that FASER-I has very little chance to explore this case, while regions to be explored by FASER-II are depicted on two plots 
of Fig.\,\ref{fig:PhotonMesonPFASER}     
for the flavor-conserving and flavor-violating sgoldstino couplings. 

\begin{figure}[!htb]  
\centerline{
\includegraphics[width=0.5\linewidth]{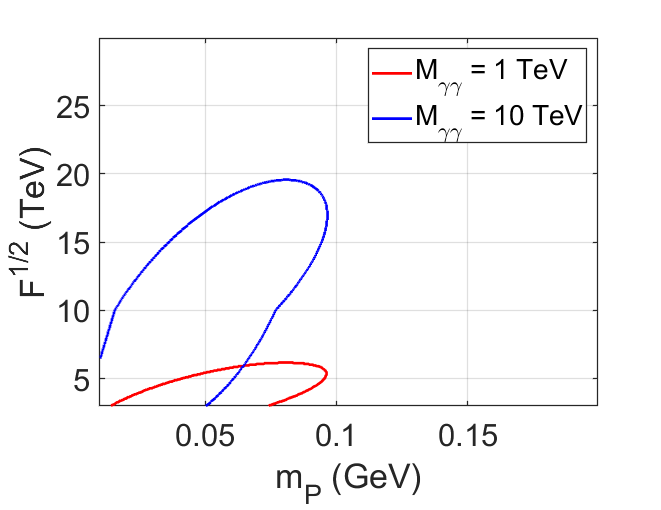}
\hskip 0.03\textwidth 
\includegraphics[width=0.5\linewidth]{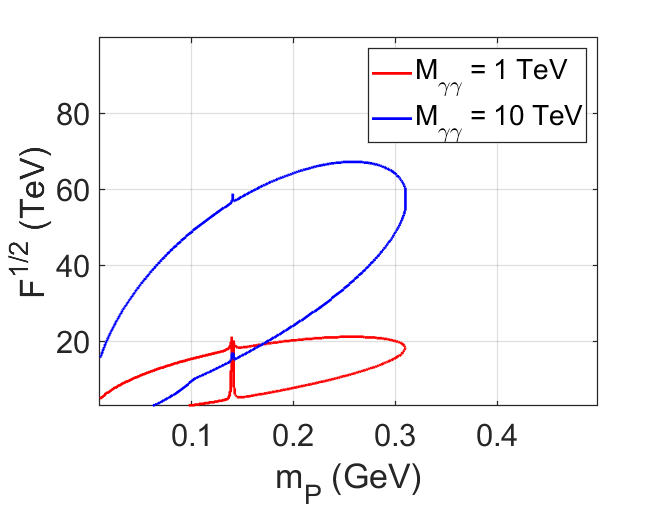} }  
\caption{Sensitivity region of FASER-I (left plot) and FASER-II (right plot) to models with sgoldstino. Pseudoscalar sgoldstinos are produced in Primakoff process and decay into a pair of photons. $M_3 = 3$ TeV. { In the regions where  $\sqrt{F}<M_{\gamma \gamma}$ we set $M_{\gamma \gamma}=\sqrt{F}$.}} 
\label{fig:PrimPFASER}
\end{figure}

The most promising source of pseudoscalar sgoldstino is Primakoff process. Primakoff process gives somewhat similar pictures for sensitivity regions on FASER. This case presented in Fig.\,\ref{fig:PrimPFASER}.

\begin{figure}[!htb]  
\centerline{
\includegraphics[width=0.5\textwidth]{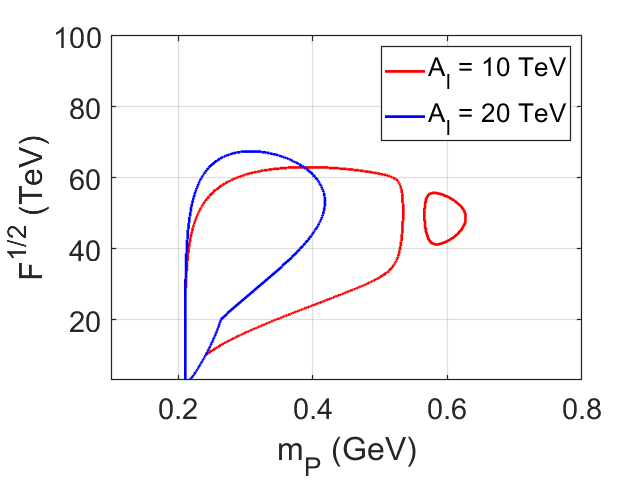}
\hskip 0.03\textwidth 
\includegraphics[width=0.5\textwidth]{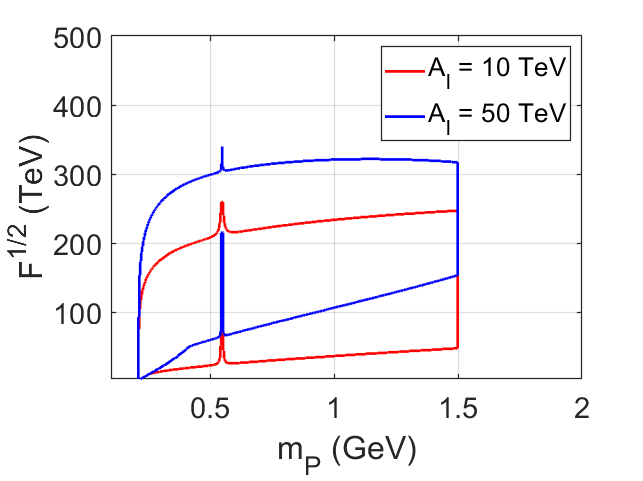}}
\caption{The regions to be tested with muon pair signature at FASER-I (left panel) and FASER-II (right panel). $M_3=3$ TeV. { In the regions where $\sqrt{F}<A_l$ we set $A_l=\sqrt{F}$.}} 
\label{fig:FusionLeptonPFASER}
\end{figure}

\paragraph{Two leptons.} Similarly to the two photon channel, this mode is also more promising for direct production mechanism than that in the scalar sgoldstino case. The experiments FASER-I and FASER-II will be able to investigate the regions presented in Fig.\,\ref{fig:FusionLeptonPFASER}
corresponding to the muon pair signature. 

\begin{figure}[!htb]  
\centerline{
\includegraphics[width=0.5\linewidth]{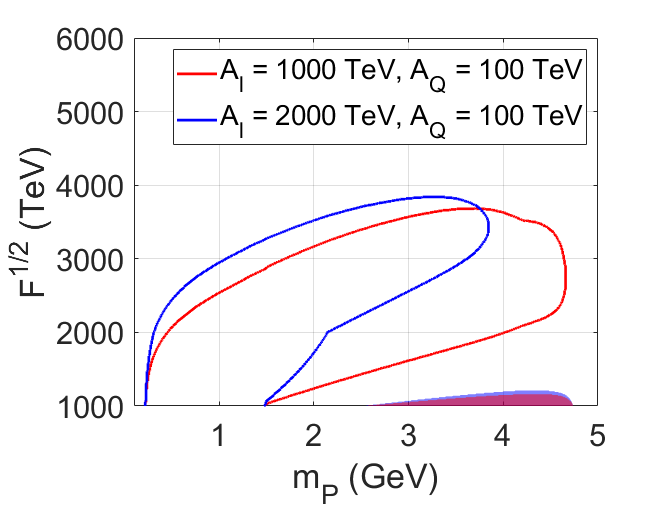}
\hskip 0.03\textwidth 
\includegraphics[width=0.5\linewidth]{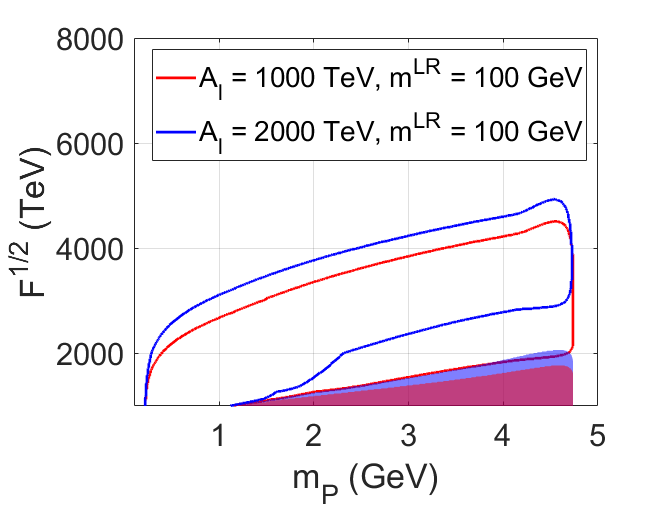} }  
\caption{Sensitivity region of FASER-II to models with sgoldstino flavor-conserving (left panel) and flavor-violating (right panel) couplings to quarks. Pseudoscalar sgoldstinos are produced in decays of beauty mesons and decay into lepton pairs. $M_3=3$ TeV. { Shaded areas correspond to restrictions from tab.\,\ref{tab:tab4}. In the regions where $\sqrt{F}<A_l$ we set $A_l=\sqrt{F}$.}} 
\label{fig:MesonLeptonPFASER}
\end{figure}

Likewise, only FASER-II is promising in testing this mode for the sgoldstino emerged in meson decays, the viable regions are presented 
in Fig.\,\ref{fig:MesonLeptonPFASER}
for both flavor-conserving (left panel) and flavor-violating (right panel) couplings. 

\begin{figure}[!htb]  
\centerline{
\includegraphics[width=0.5\textwidth]{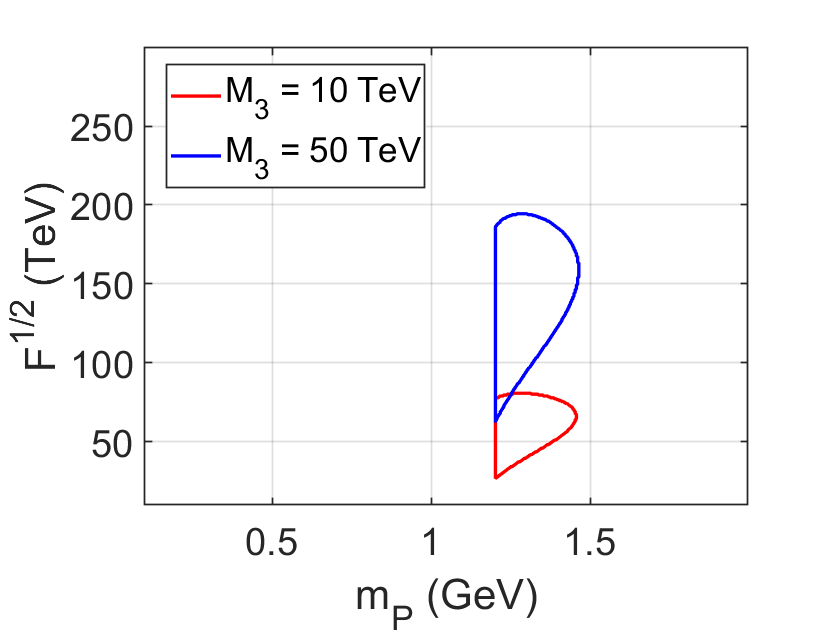} 
\hskip 0.03\textwidth 
\includegraphics[width=0.5\textwidth]{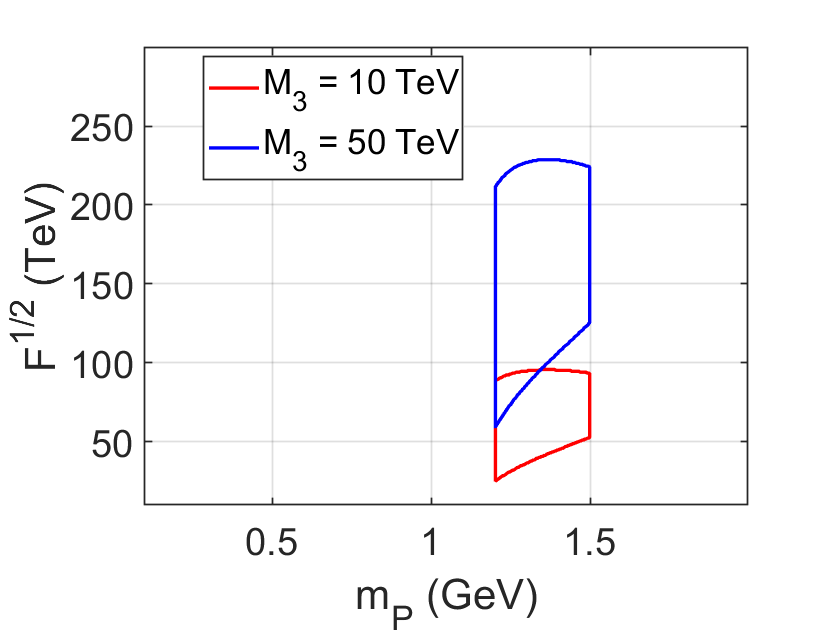} 
}
\caption{Sensitivity regions of FASER-I to the model parameters, where pseudoscalar sgoldstinos are produced directly and then decay inside the detector volume into mesons. When only charged mesons are used as a signature, the regions are presented on left panel. The right panel shows the region, when both charged and neutral mesons are adopted as sgoldstino signatures. $A_Q=100$ GeV. { In the regions where $\sqrt{F}<M_3$ we set $M_3=\sqrt{F}$.}} 
\label{MesonFissionPFASER1}
\end{figure}

\begin{figure}[!htb]  
\centerline{
\includegraphics[width=0.5\textwidth]{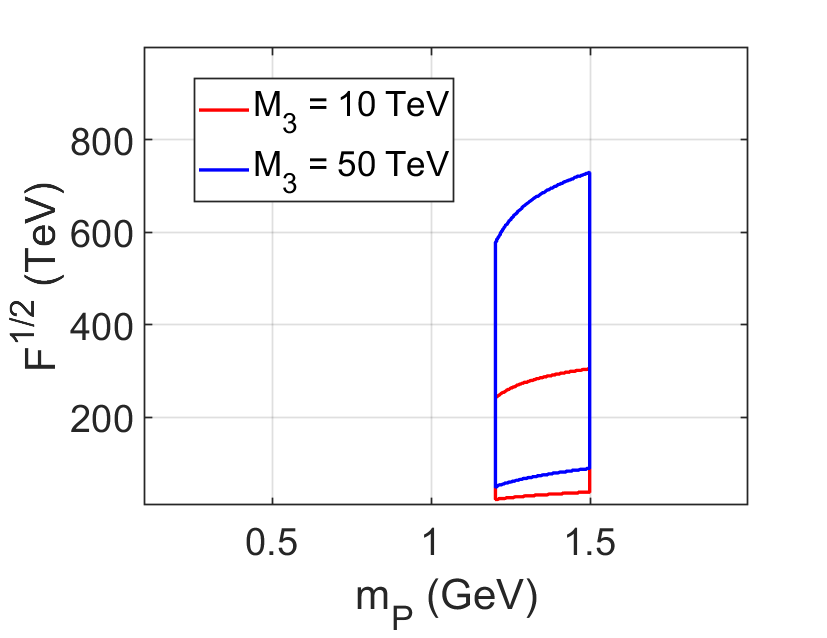} 
\hskip 0.03\textwidth 
\includegraphics[width=0.5\textwidth]{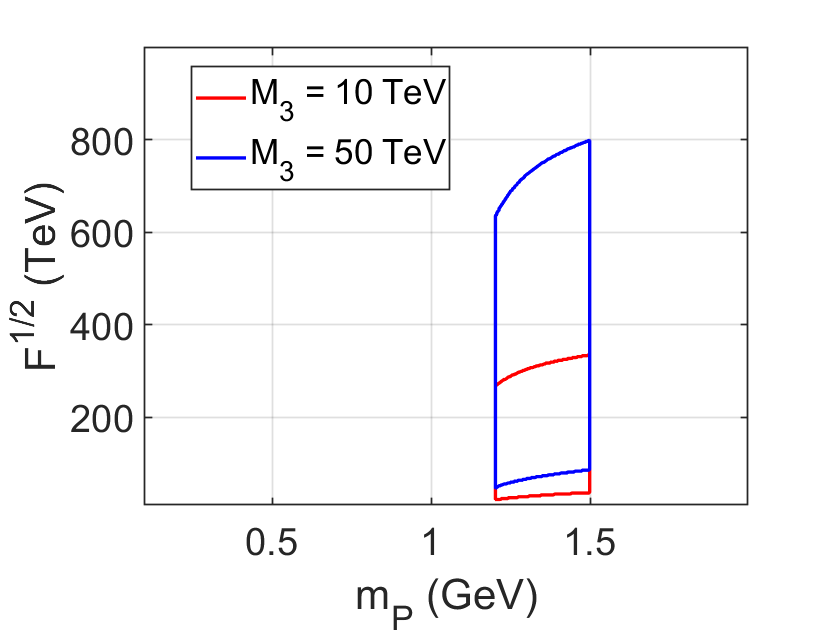} 
}
\caption{Sensitivity regions of FASER-II to the model parameters, where pseudoscalar sgoldstinos are produced directly and then decay inside the detector volume into mesons. When only charged mesons are used as a signature, the regions are presented on left panel. The right panel shows the region, when both charged and neutral mesons are exploited as sgoldstino signatures. $A_Q=100$ GeV. { In the regions where $\sqrt{F}<M_3$ we set $M_3=\sqrt{F}$.}} 
\label{MesonFissionPFASER2}
\end{figure}

\begin{figure}[!htb]  
\centerline{
\includegraphics[width=0.5\textwidth]{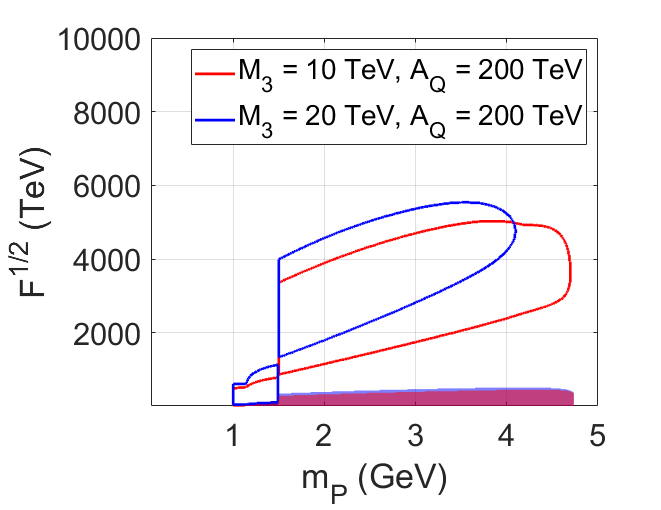}  
\hskip 0.03\textwidth 
\includegraphics[width=0.5\textwidth]{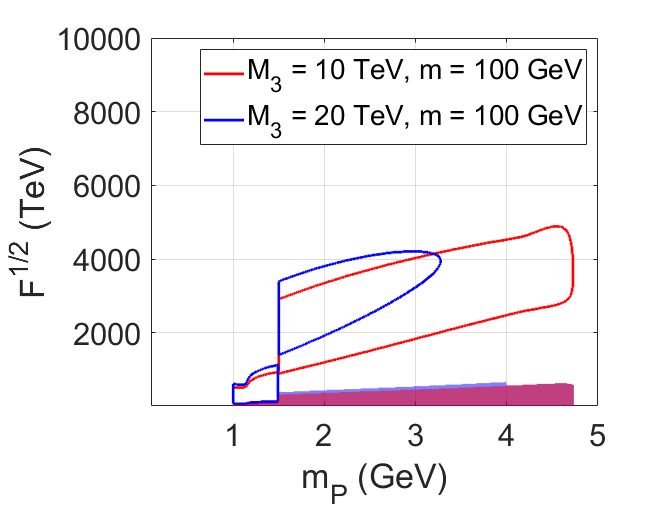} 
    }
\caption{FASER-II sensitivity to light pseudoscalar sgoldstinos produced in B-meson decays via flavor-conserving couplings (left panel) and flavor-violating couplings with $A_Q = 100$ GeV(right panel). Sgoldstino decays into a gluon pair inside the detector volume. {Shaded areas correspond to restrictions from tab.\,\ref{tab:tab4}. In the regions where $\sqrt{F}<M_3$ we set $M_3=\sqrt{F}$ and similarly for $A_Q$.} } 
\label{fig:MesonGluonsPFASER}
\end{figure}

\paragraph{Mesons.}
Both direct production and decay rates grow with gluino mass $M_3$. Pseudoscalar sgoldstinos lighter than 1.5\,GeV decays into three mesons, hence the hadronic decay mode is suppressed with respect to the scalar sgoldstino case, and so the pseudoscalar sgoldstino decay length is larger. For heavier sgoldstinos the scalar and pseudoscalar cases are the same. Both FASER-I and FASER-II can probe small range of sgoldstino masses as shown in Fig.\,\ref{MesonFissionPFASER1} 
and Fig.\,\ref{MesonFissionPFASER2} 
respectively. Only FASER-II can probe wider range of sgoldstino masses as shown in  Fig.\,\ref{fig:MesonGluonsPFASER},
where sgoldstino decays into gluons.

In conclusion we see that in supersymmetric models with light pseudoscalar sgoldstino, FASER can test vast regions of the model parameter space, which are different but partly overlap with those in case of light scalar sgoldstino.  
We illustrated this statement above with several plots. Our numerical study allows for outlining these regions.  

When sgoldstinos decay mostly to photons, FASER-I can test the following region
\begin{equation}
\begin{split}
    & \sqrt{F} < 2000 \, \text{TeV} \\
    & m_S < 1.5 \, \text{GeV} \\
    & M_{\gamma\gamma}< \sqrt{F} \\
    & M_3< \sqrt{F}
\end{split}
\end{equation}
and FASER-II
\begin{equation}
\begin{split}
    & \sqrt{F} < 170 \cdot 10^3 \, \text{TeV} \\
    & m_S < 4.8 \, \text{GeV} \\
    & m_{ij}^{LR}<100 \, \text{GeV} \\
    & M_{\gamma\gamma}< \sqrt{F} \\
    & A_Q< \sqrt{F}
\end{split}
\end{equation}
If case of dominant lepton mode, FASER-I can test the following region
\begin{equation}
\begin{split}
    & \sqrt{F} < 1400 \, \text{TeV} \\
    & 2m_\mu < m_S < 1.5 \, \text{GeV} \\
    & A_l< \sqrt{F}
\end{split}
\end{equation}
and FASER-II
\begin{equation}
\begin{split}
    & \sqrt{F} < 50 \cdot 10^3 \, \text{TeV} \\
    & 2m_\mu  < m_S < 4.8 \, \text{GeV} \\
    & m_{ij}^{LR}<100 \, \text{GeV} \\
    & A_l< \sqrt{F} \\
    & A_Q< \sqrt{F}
\end{split}
\end{equation}
In sgoldstino models, where hadronic decay modes dominate, FASER-I can test the following region
\begin{equation}
\begin{split}
    & \sqrt{F} < 1300 \, \text{TeV} \\
    & 1.2 \, \text{GeV} < m_S < 1.5 \, \text{GeV} \\
    & M_3< \sqrt{F} \\
    & A_Q < \sqrt{F}
\end{split}
\end{equation}
while FASER-II 
\begin{equation}
\begin{split}
    & \sqrt{F} < 200 \cdot 10^3 \, \text{TeV} \\
    & 3m_\pi < m_S < 4.8 \, \text{GeV} \\
    & m_{ij}^{LR}<100 \, \text{GeV} \\
    & M_3< \sqrt{F} \\
    & A_Q< \sqrt{F} \,.
\end{split}
\end{equation}

\section{Conclusions}
\label{Sec:conclusion}
To summarize, we investigate FASER prospects in searches for light  ($0.2-5$\,GeV) scalar and pseudoscalar sgoldstinos in low energy supersymmetric extensions of the Standard Model of particle physics. We consider both first (FASER-I) and second (FASER-II) stages of FASER operation, and considered three viable signatures of sgoldstino decays inside the FASER detector: emerged from a single point  couple of photons, couple of muons or light mesons. In proton-proton scatterings sgoldstino can be produced directly and in the subsequent decays of produced mesons. In the last case both flavor-conserving and  { flavor-violating} sgoldstino couplings to quarks may contribute. 

We take the model with all SM superpartners to be sufficiently heavy, above the kinematical threshold of direct production at LHC, and hence irrelevant for our study. However, sgoldstino couplings to the SM fields are proportional to the corresponding soft supersymmetry breaking parameters, and thus affect sgoldstino phenomenology. All sgoldstino couplings are inversely proportional to the supersymmetry breaking parameter $F$ in the entire model, which is of order of the squared energy scale of supersymmetry breaking. The brightest and unique option related to searches for light sgoldstino is related to direct measurement of its couplings and inference of the value of the supersymmetry breaking scale.   

We illustrate the FASER  (both stages) perspectives for each decay signature and each production mechanism by choosing particular sets of the soft terms and depicting a region of sgoldstino mass and supersymmetry breaking scale $\sqrt{F}$, where one can expect 3 signal events to be observed inside the FASER detector. The results presented in numerous Figures show that FASER-II is much more promising in this respect, however the FASER-I has also a large room for investigation of this model. We also roughly outlined the region in the whole model parameter space (soft terms, sgoldstino mass and supersymmetry breaking scale), where FASER can search the light sgoldstinos in each of the cases above.  Broadly speaking, we conclude that with the FASER one can probe the models with light sgoldstinos having supersymmetry breaking scale $\sqrt{F}\lsim 1500-5000$~TeV at the first stage of experiment and about 1--2 orders of magnitude higher at the second one.

There are several final remarks. {The first} concerns the sgoldstino flavor-violating couplings to quarks, which may contribute to sgoldstino production in the meson decays. These couplings are presently constrained from direct searches of the {rare meson} decays with missing energy (sgoldstino flies away) and with pair of SM particles (emerged from sgoldstino decays). We took these constraints into account. At the same time, since these couplings are proportional to the off-diagonal entries in squark mass squared matrices, they are also limited from rare processes (e.g. Flavor Changing Neutral Currents, FCNC) which may receive contributions from virtual squarks. The latter contributions are proportional to the off-diagonal entries, but inversely proportional to squark squared masses. In our case of very heavy squarks these contributions are suppressed, and so the off-diagonal elements are less constrained, allowing them to be of the same order as the diagonal elements for sufficiently heavy squarks. Thus, bounds from FCNC are always obeyed in our study. When outlying the regions in model parameter space to be tested at FASER we always limit the off diagonal terms to be less than $(100\,\text{GeV})^2$. With heavy squarks this bound may be noticeably relaxed, as far as direct limits from searches for sgoldstino in meson decays are satisfied. 

{The second} remark is related to theoretical uncertainties pertinent to the predictions made in this study. Many estimates presented here concern production and/or decay of new light scalar particle which are mediated by strong interactions. As we already mentioned in Sec.\,\ref{Subsec:decays}, the uncertainties in calculations of the sgoldstino decay width to hadrons are up to 2 orders of magnitude. Likewise, there are uncertainties in our estimates of sgoldstino direct production in case of small sgoldstino masses, less than $1-1.5$\,GeV. For the present study we used a simple extrapolation of DIS results from the case of heavy sgoldstino masses. And although this cross section does scale as $M_3^2/F^2$ one should be careful when obtaining actual bounds from future experimental results due to the inherent  uncertainties in this procedure.

\vskip 0.5cm 

We thank J.\ Feng  and A.\ Arbuzov for valuable comments. The work is supported by the Russian Science Foundation RSF grant 21-12-00379. 
\bibliographystyle{apsrev4-1}
%
\bibliography{refs}

\end{document}